\documentclass[10pt,journal,compsoc]{IEEEtran}

\ifCLASSOPTIONcompsoc
  \usepackage[nocompress]{cite}
\else
  \usepackage{cite}
\fi

\usepackage{tabu}
\usepackage{bm}
\usepackage{booktabs} 
\usepackage{makecell} 
\usepackage{siunitx, mhchem}

\usepackage{comment}
\usepackage{array}
\usepackage{multirow,afterpage}
\usepackage{wrapfig}
\usepackage[abs,unit=1mm]{overpic}

\usepackage{enumitem}
\usepackage[dvipsnames]{xcolor}

\usepackage{wrapfig,lipsum,booktabs}
\usepackage{amssymb}

\usepackage[bookmarks,backref=true,linkcolor=black]{hyperref} 
\hypersetup{
  pdfauthor = {},
  pdftitle = {},
  pdfsubject = {},
  pdfkeywords = {},
  colorlinks=true,
  linkcolor= black,
  citecolor= black,
  pageanchor=true,
  urlcolor = black,
  plainpages = false,
  linktocpage
}

\renewcommand{\baselinestretch}{0.965}

\begin{document}
\title{Mechanics-Aware Modeling\\of Cloth Appearance}


\author{Zahra Montazeri, Chang Xiao, Yun (Raymond) Fei, Changxi Zheng, and Shuang Zhao}

\author{Zahra Montazeri, Chang Xiao, Yun (Raymond) Fei, Changxi Zheng, and Shuang Zhao
\IEEEcompsocitemizethanks{\IEEEcompsocthanksitem Z. Montazeri and S. Zhao was with the Department
of Computer Science, University of California, Irvine,
CA, 92697.\protect\\
E-mail: zmontaze@uci.edu
\IEEEcompsocthanksitem C. Xiao, Y. (R.) Fei and C. Zheng are with Columbia University, New York, NY, 10027}
\thanks{Manuscript received May 12, 2019; revised May 12, 2019.}}

\markboth{IEEE Transactions on Visualization and Computer Graphics}%
{Montazeri \MakeLowercase{\textit{et al.}}: Mechanics-Aware Modeling of Cloth Appearance}

\iffalse
\newcommand{\sz}[1]{{\color{orange}[SZ: #1]}}
\newcommand{\zahra}[1]{{\color{blue}[Zahra: #1]}}
\newcommand{\fixme}[1]{{\color{red}[#1]}}
\newcommand{\todo}[1]{\marginpar{\Large {\color{red} $\spadesuit$}} {\bf \color{red} [#1]}}
\newcommand{\rev}[1]{{\color{blue}#1}}
\else
\newcommand{\sz}[1]{}
\newcommand{\zahra}[1]{}
\newcommand{\fixme}[1]{}
\newcommand{\todo}[1]{}
\newcommand{\rev}[1]{#1}
\fi

\newcommand{\E}{\mathrm{e}}
\newcommand{\bx}{\bm{x}}
\newcommand{\by}{\bm{y}}
\newcommand{\bn}{\bm{n}}
\newcommand{\bb}{\bm{b}}
\newcommand{\bt}{\bm{t}}
\newcommand{\bR}{\bm{R}}
\newcommand{\bT}{\bm{T}}
\newcommand{\bF}{\bm{F}}
\newcommand{\calt}{\mathcal{T}}
\newcommand{\bom}{\bm{\omega}}
\newcommand{\bal}{\bm{\alpha}}
\newcommand{\bsm}{\bm{m}}
\newcommand{\bh}{\bm{h}}
\newcommand{\BO}{\mathbb{B}_0}
\newcommand{\intd}{\,\mathrm{d}}
\newcommand{\Real}{\mathbb{R}}
\newcommand{\Sph}{\mathbb{S}^2}
\newcommand{\pr}{\mathbb{P}}

\newcommand{\dg}{\bF}
\newcommand{\dglocal}{\dg^\mathrm{local}}
\newcommand{\fiberT}{\bT}
\newcommand{\fiberTlocal}{\fiberT^\mathrm{local}}
\newcommand{\regnet}{\mathcal{N}}

\newcommand\relphantom[1]{\mathrel{\phantom{#1}}}
\newcommand{\argmin}{\operatornamewithlimits{argmin}}

\newcommand{\figref}[1]{Figure~\ref{fig:#1}}
\newcommand{\tabref}[1]{Table~\ref{tab:#1}}
\newcommand{\eqnref}[1]{Eq.~\eqref{eq:#1}}
\newcommand{\secref}[1]{\S\ref{sec:#1}}
\newcommand{\appref}[1]{Appendix~\ref{sec:#1}}
\newcommand{\eq}[1]{\eqref{eq:#1}}

\newlength\savedwidth
\newcommand{\whline}[1]{\noalign{\global\savedwidth\arrayrulewidth
                              \global\arrayrulewidth #1} %
                     \hline
                     \noalign{\global\arrayrulewidth\savedwidth}}

\hypersetup{draft}

\IEEEtitleabstractindextext{%
\begin{abstract}
	Micro-appearance models have brought unprecedented fidelity and details to cloth rendering.
	Yet, these models neglect fabric mechanics: when a piece of cloth interacts with the environment, its yarn and fiber arrangement usually changes in response to external contact and tension forces.
	Since subtle changes of a fabric's microstructures can greatly affect its macroscopic appearance, mechanics-driven appearance variation of fabrics has been a phenomenon that remains to be captured.
	
	We introduce a mechanics-aware model that adapts the microstructures of cloth yarns in a physics-based manner.
	Our technique works on two distinct physical scales: using physics-based simulations of individual yarns, we capture the rearrangement of yarn-level structures in response to external forces.
	These yarn structures are further enriched to obtain appearance-driving fiber-level details.
	The cross-scale enrichment is made practical through a new parameter fitting algorithm for simulation, an augmented procedural yarn model coupled with a custom-design regression neural network.
	We train the network using a dataset generated by joint simulations at both the yarn and the fiber levels.
	Through several examples, we demonstrate that our model is capable of synthesizing photorealistic cloth appearance in a 
	mechanically plausible way.
\end{abstract}

}







\maketitle

\IEEEdisplaynontitleabstractindextext

%
\IEEEpeerreviewmaketitle


\begin{figure*}[t]
	\centering
	\makebox[\textwidth][c]{\includegraphics[width=1.18\textwidth]{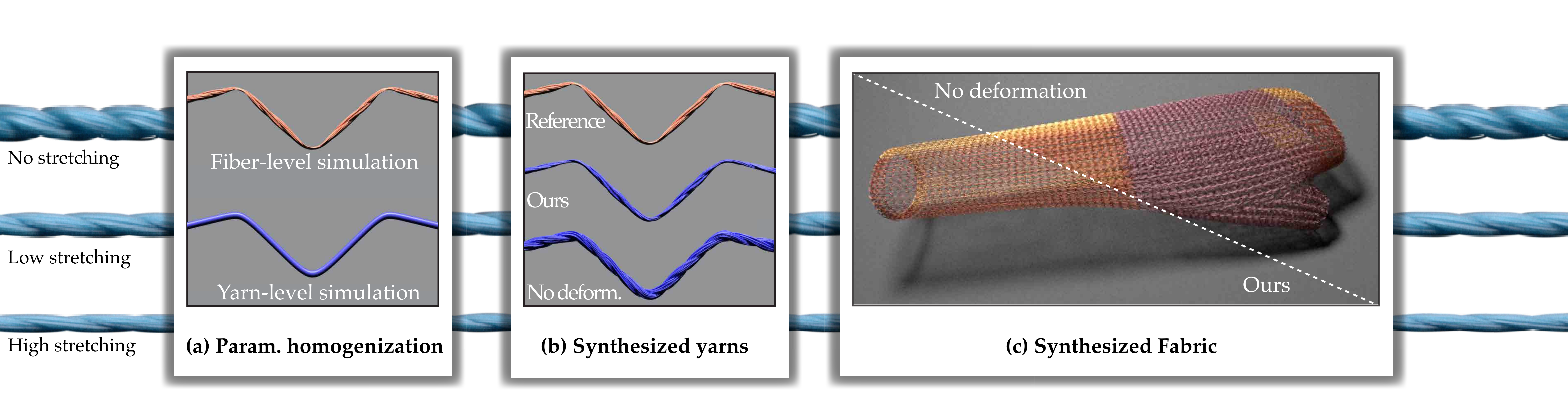}}%
	\caption{\label{fig:teaser}
		We introduce a first-of-a-kind technique to model cloth appearance not only micro-geometrically but mechanically.
		(a)~Our technique simulates the deformation of fibers at the yarn-level by leveraging a novel parameter homogenization process so that centerlines of a yarn simulated at the yarn and the fiber levels match.
		(b)~By procedurally generating fibers around simulated yarn centerlines and deforming those fibers based on yarn-level external forces, our method synthesizes fiber microstructures that closely match reference fiber-level simulations.
		(c)~Accurate reproduction of fiber-level micro-geometries is crucial as they greatly affect a fabric's overall appearance.
	}
\end{figure*}

\IEEEraisesectionheading{\section{Introduction}\label{sec:intro}}
It is a universal phenomenon that 
a material's microstructure affects its macroscopic appearance profoundly.  
In light of this, computer graphics has made
significant advances toward capturing realistic object appearance by modeling its
microstructures (e.g.,
\cite{Cook:1982:RMC,Jakob:2014:DSM,Dong:2015:PAM,Liu:2016:SST}).
Motivated by its utter prominence in virtual senses, cloth appearance has also
been modeled microscopically. 
State-of-the-art \emph{micro-appearance}
models~\cite{Zhao:2011:BVA,Khungurn:2015:MRF,Zhao:2016:FPY} can produce cloth
renderings with stunning richness and details, by accounting for the fabric's 
\emph{micro-geometric structures} (i.e., the arrangement of fibers and yarns). 

\begin{figure}[b]
	\centering
	\addtolength{\tabcolsep}{-3.5pt}
	\begin{tabular}{cc}
		\begin{overpic}[width=0.23\textwidth]{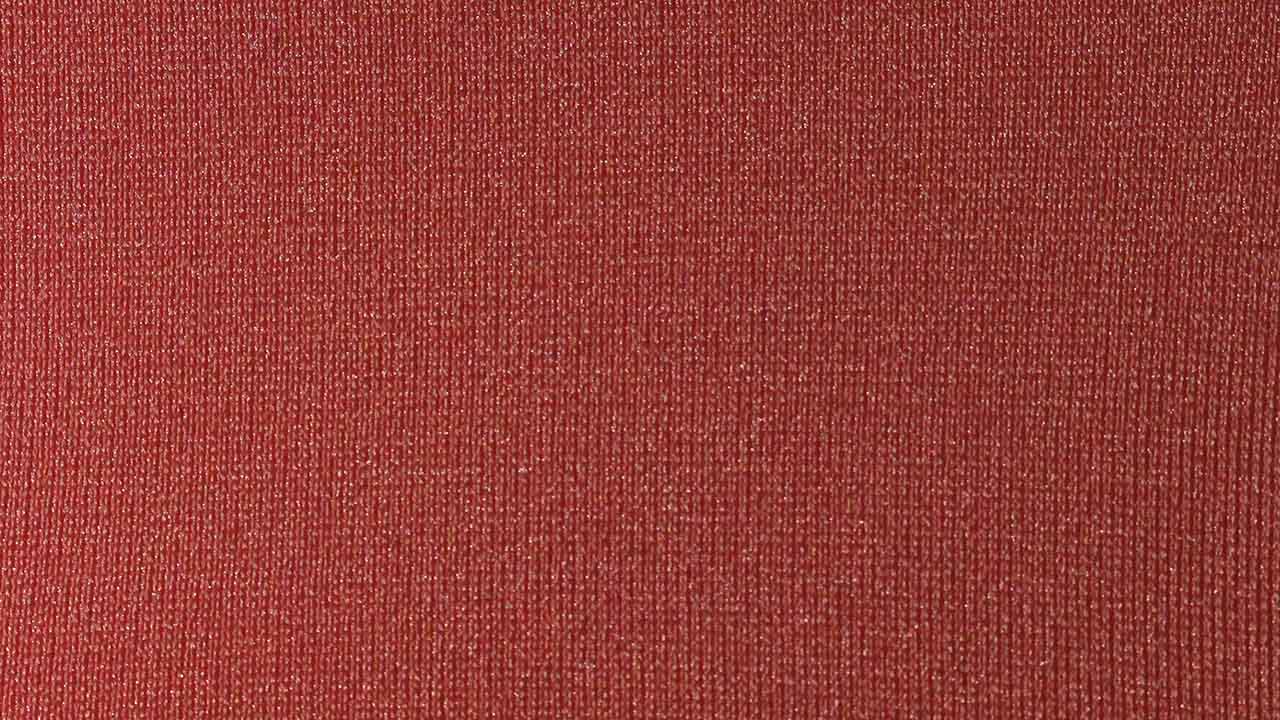}
                    \put(2, 3){\small \color{white} \bfseries Rest}
		\end{overpic}
		&
		\begin{overpic}[width=0.23\textwidth]{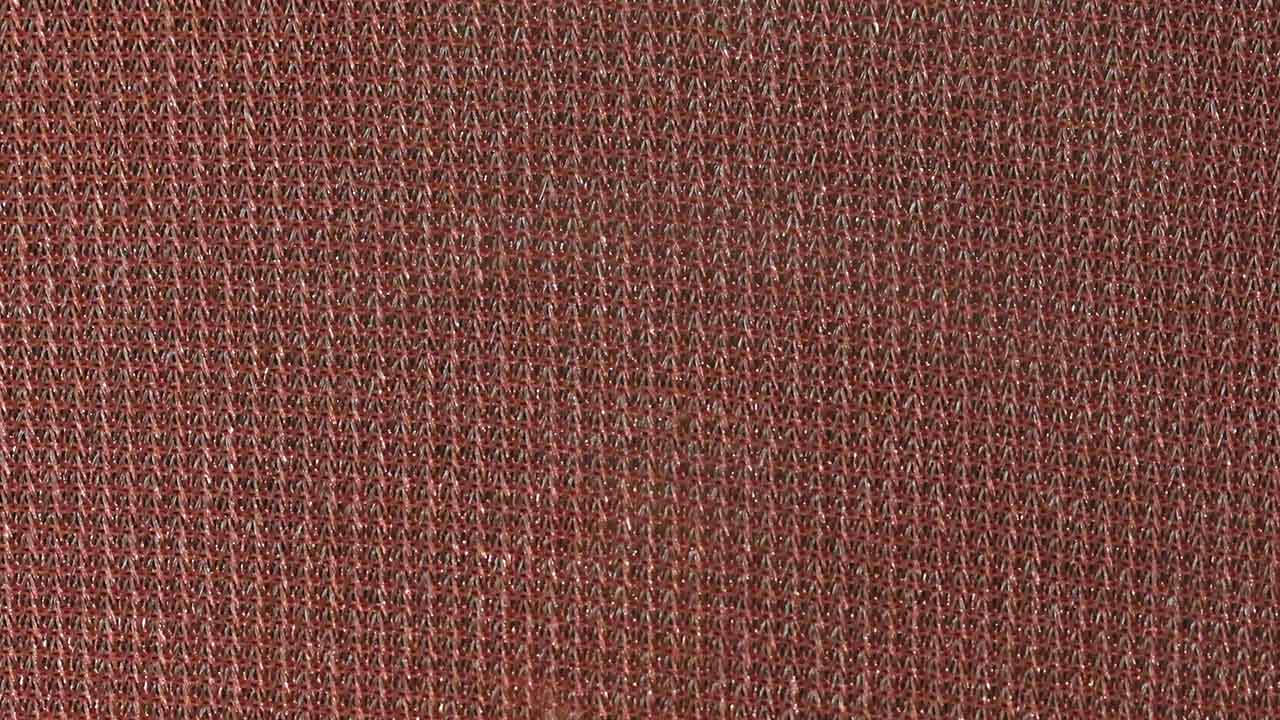}
                    \put(2, 3){\small \color{white} \bfseries Stretched}
		\end{overpic}
	\end{tabular}
    \caption{\label{fig:motivating_example}
    	\textbf{Appearance change of a real fabric} driven by external stretching.
    	When a real piece of cloth is stretched, its internal yarn and fiber structures can change drastically, leading to significantly different overall appearance.
    	In this example, we show photographs of a nylon fabric under rest and stretched states.
    }
\end{figure}

We argue that the microstructures of cloth should be modeled not only geometrically but also \emph{mechanically}.
This is because the appearance-driving arrangement of a fabric's constituent fibers and yarns is by no means fixed. 
For instance, a simple stretch of a piece of fabric may thin its yarns, making the fabric more see-through and changing its shininess (see Figure~\ref{fig:motivating_example}).
Even without external distortion, the interaction between individual yarns of a cloth can change yarn shapes 
and cause specific texture patterns to emerge~\cite{Kaldor2008SKC,Cirio2014YSW}.
The mechanical response---the interplay between small-scale filament structures and their material properties---plays a significant role in a fabric's appearance.

From this vantage point, we introduce a cloth appearance model that
incorporates mechanical responses of a fabric's microstructures for better
capturing the fabric appearance and its changes under external forces.

A seemingly straightforward idea is to rely on physics-based simulation 
to reconfigure the structures of yarns and fibers. 
However, executing this idea needs to address 
a significant computational challenge.
On the one hand, it has been shown that both the structure of fibers (i.e., how
the individual filaments form a yarn) and the structure of yarns (i.e., how the
individual yarns are interwoven) are instrumental for fabric
appearance~\cite{Zhao:2011:BVA,Khungurn:2015:MRF,schroder2015image,Zhao:2016:FPY,Carlos2017}.
On the other hand, while there exist physics-based models that simulate 
individual yarns of a cloth~\cite{Kaldor2010EYC,Cirio2014YSW}, simulating
individual fibers is intractable, because 
even for a sizable cloth, a myriad of fibers must be included.

We tackle this challenge through a \emph{bi-scale} approach: we simulate the
cloth only at the \emph{yarn level} and synthesize the \emph{fiber-level}
details in a mechanics-aware manner 
during post-processing.
The synthesis of fibers is enabled by augmenting the procedural yarn
model~\cite{schroder2015image,Zhao:2016:FPY} with spatially varying parameters
derived from the simulated yarn states.
These parameters are used to deform the procedurally generated fibers
in the way that reflect the fabric's mechanical response.
Since the relationship between
the fiber-deforming parameters and the simulated yarn-level states is highly
complex, we leverage a custom-design regression neural network to learn this
relation.

The training dataset of our regression network is generated by running physics-based
simulations at both the yarn level (where a cloth yarn is modeled as a single
curve) and the fiber level (where a yarn is depicted as a bundle of fiber
curves). To make the training simulations tractable, we design the neural
network such that they involve only a small yarn segment. 
Furthermore, to ensure the consistency between 
the simulations at both scales, we develop a new parameter fitting algorithm
that automatically determines the (homogenized) material properties at the yarn
level using those from the fiber level.

At runtime, our method simulates the microstructures of a fabric at the yarn level.
It then utilizes procedural yarn modeling coupled with the pre-trained
regression network to generate the fabric's fiber-level geometries
and adapt them to the simulated yarn states. In this way, we are able to produce diverse and
physically plausible fabric appearance (Figure~\ref{fig:teaser}).  

To our knowledge, this is the first mechanics-aware technique for modeling the appearance of cloth (or any material if that matters): existing techniques offering fiber-level details either rely
on measured fiber geometries that are virtually impossible to 
animate~\cite{Zhao:2011:BVA,Khungurn:2015:MRF}, or use procedurally generated
fibers with fiber and yarn mechanics completely
neglected~\cite{schroder2015image,Zhao:2016:FPY,Luan:2017:FOP}.
Our technique is also independent of the underlying cloth simulation method:
although we choose to use a state-of-the-art offline approach for maximized
accuracy, other yarn-level simulation approach can be readily used as well.
Our major technical contributions include:
\begin{itemize}
\item A \emph{parameter homogenization} technique that fits yarn-level mechanical parameters so that yarn centerlines match between fiber-level and yarn-level simulations.
\item A \emph{custom-design regression neural network} that learns the mapping from simulated yarn-level states to fiber-deforming parameters.
\end{itemize}

We validate our method by comparing fiber-level simulations and photographs (Figures~\ref{fig:result_single} and \ref{fig:photo_comp}).
To demonstrate the use of our method, we apply it to a few fabrics under various mechanical deformations, and show their appearance changes in response to the interaction with the environment (Figures~\ref{fig:teaser}, \ref{fig:result_knitted}, and \ref{fig:result_woven}).


\section{Related Work}\label{sec:related}

\paragraph*{\em Cloth appearance models}
Modeling and reproducing the appearance of a cloth has been an active research
area in computer graphics for decades. Traditionally, cloth are modeled as 2D
thin sheets with the appearance described using
generic~\cite{Westin:1992:PRF,Ashikmin:2000:MBG,Wu:2011:PIB} or specialized surface
reflectance models~\cite{Adabala:2003:VWC,Irawan:2012:SRW,Sadeghi:2013:PMA}.
These models offer adequate quality for reproducing cloth appearance when
viewed from a distance where individual cloth yarns are barely visible.
\rev{On the other hand}, they lack \rev{the fidelity and} details to produce plausible close-up
renderings and the generality to capture fabrics' diverse visual
effects (such as strong textures and grazing-angle highlights)
usually arising from their characteristic microstructures. 

\begin{figure*}[t]
	\centering
	\includegraphics[width=0.99\textwidth]{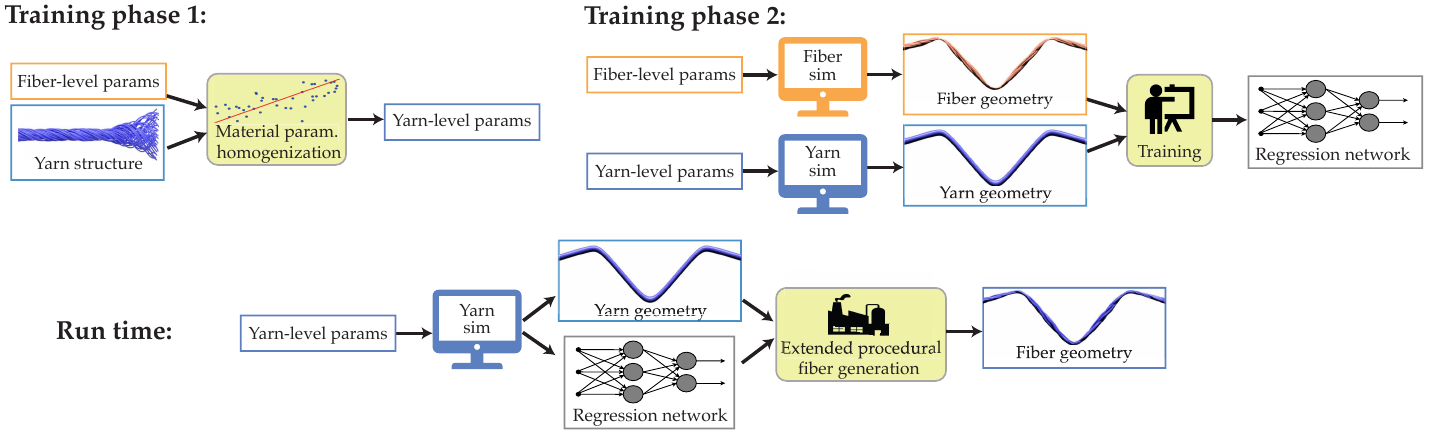}
	\caption{\label{fig:overview}
		\textbf{Method overview.}
		\textbf{(top)} In the preprocessing phase, our method has two sub-steps: homogenizing the material parameters for yarn-level simulation (left) and training the regression neural network (right).
		\textbf{(bottom)} At runtime, we capture the rearrangement of yarn-level structures in a fabric, and enrich it with fiber-level geometries with the neural network. The fiber-level geometries then determine the macroscopic appearance of the fabric.
	}
\end{figure*}


Recently, a family of micro-appearance models has been introduced 
to model fabric appearance~\cite{Zhao:2011:BVA,Zhao:2012:SSP,schroder2015image,Khungurn:2015:MRF,Zhao:2016:FPY,Wu2017,Carlos2017,Luan:2017:FOP,Khungurn:2017:FRF}.
Unlike earlier methods, these techniques explicitly model fabrics'
fiber-level microstructures using high-resolution volumes or fiber meshes.
Since these microstructures largely affect a fabric's overall appearance,
micro-appearance models enjoy the unprecedented details and the generality to capture the appearance of a
wide variety of fabrics.  

Unfortunately, these models are mechanically agnostic: many of them rely on
measured microstructures that cannot be mechanically deformed at all~\cite{Zhao:2011:BVA,Zhao:2012:SSP,Khungurn:2015:MRF,Khungurn:2017:FRF};
others use specialized procedures that completely neglect fiber
mechanics~\cite{schroder2015image,Zhao:2016:FPY,Wu2017,Luan:2017:FOP},
producing inaccurate overall appearance (e.g., see
Figure~\ref{fig:motivating_example}). In short, these models are significantly
limited when modeling animated fabrics.

\paragraph*{\em Cloth and yarn simulation}
Computer graphics has also seen a long history of simulating cloth and yarns. A
thorough review can be found in the survey~\cite{thomaszewski2007advanced}. 
Oftentimes, a piece of cloth is modeled as a 2D surface, discretized
with a triangular or quadrilateral mesh~\cite{Baraff:1998:LSC,grinspun2003discrete}. 
Since a real cloth is woven by a set of fiber bundles (or \emph{yarns}), 
a more faithful method is to model those yarns directly.
The first of such models was introduced by Kaldor et al.~\cite{Kaldor2008SKC}, 
and later accelerated in~\cite{Kaldor2010EYC}.

The main challenge of simulating at yarn level is the high cost of resolving contacts. 
To address this issue, Cirio et al.~\cite{Cirio2014YSW} introduced sliding
constraints to efficiently treat inter-yarn contacts. 
Our work is built on a Lagrangian/Eulerian approach of resolving contacts~\cite{jiang2017anisotropic}.
Extending the material point method~\cite{sulsky1994particle}, this approach represents yarns
as polylines embedded in a background Eulerian grid, and 
resolves collisions by modeling them as volumetric deformations.
In the same framework, this method can also simulate thin shells, while ignoring bending and twisting forces.
This limitation is addressed recently for mesh-based thin shells~\cite{guo2018mpm}.
We address this limitation for simulating individual yarns by incorporating the elastic rod model~\cite{bergou2010discrete},
similar to the recent work by Fei et al.~\cite{fei2018mms}.



Despite of these advances, simulating individual fibers in a sizable cloth remains intractable,
because an excessively large number of fibers is needed.
While Jiang et al.~\cite{jiang2017anisotropic} managed to simulate multiple ``threads'' per yarn in 
their examples, the number of threads is still much smaller than the typical number of fibers in a real yarn.
In this work, we sidestep the challenge of simulating fibers fully. Our simulation still performs at the yarn level, but 
we design a deep neural network to ``upsample'' the yarn-level simulation results to obtain fiber-level details.


\paragraph*{\em Cloth rendering vs. simulation} 
Historically, the research on cloth appearance modeling and physics-based
simulation advance largely in parallel, probably because the two lines of research
have been traditionally considered from different physical perspectives (optics vs. mechanics).
Although cloth simulation results are often demonstrated with realistic
rendering, it is very rare, if not at all, that a cloth appearance model has
benefited from physics-based simulation. In this work, we show that the two seemingly 
separated types of models can work in tandem, yielding an improved cloth appearance model. 

\section{Method Overview}\label{sec:overview}
The structure of fabrics is complex at multiple scales.
As a whole, a fabric consists of thousands of threads, or \emph{yarns}, combined via manufacturing techniques like weaving and knitting.
Each yarn is, in turn, comprised of tens to hundreds of micron-radius filaments, or \emph{fibers}.
Since a fabric's yarn and fiber arrangements significantly affects its overall appearance, our goal is to capture cloth's mechanical response down to the fiber level.

To this end, we leverage physics-based simulation to capture a fabric's yarn-
and fiber-level mechanics.  Unfortunately, simulating the dynamics of all fibers in an
entire fabric is generally intractable. Instead, we propose to simulate at yarn level 
(at runtime), and then procedurally
generate fibers around simulated yarns in a mechanics-aware manner to complete the
appearance model.

To enable mechanics-aware generation of fiber structures, our method consists of three major components:
\begin{itemize}
\item{\textit{Training.}}
In our training phase, we simulate a single yarn under a range of conditions at both the yarn and fiber levels.
For each pair of simulation results, we compare the simulated fiber curves and procedurally generated ones (guided by the simulated yarn centerline).
The discrepancy between the two is learned by a custom-design regression network, one that maps the simulated yarn-level states (i.e., mechanical forces) to the discrepancy-compensating fiber deformations.

\item{\textit{Material parameter homogenization.}}
The training step demands consistency between the yarn- and
fiber-level simulations.  Therefore, given the material properties of
individual fibers, we determine their yarn-level counterparts to
ensure the consistency between yarn centerlines simulated at both levels (under identical
initial conditions).  This is achieved by fitting (or
homogenizing) yarn-level material parameters using fiber-level simulations.

\item{\textit{Runtime.}}
At runtime, we simulate the dynamics of full fabrics only at the yarn level.
With individual yarn centerlines and forces along them determined by the simulation, we procedurally generate fibers following these centerlines and use the trained regression network to further adjust the fibers so that they respond to yarn-level forces such as stretching and compression properly.
\end{itemize}
The pipeline described above is also summarized in \figref{overview}.
In the following sections, we detail our training stage in \S\ref{sec:training2}, parameter fitting in \S\ref{sec:training1}, and runtime phase in \S\ref{sec:runtime}.

\section{Training Phase}\label{sec:training2}
The objective of our training phase is to acquire a mapping between yarn-level
deformation and the corresponding fiber reconfiguration.
This mapping will be used at run-time to procedurally generate fiber structures in a mechanics-aware manner.
To learn the mapping, we simulate a single yarn under a few predetermined
configurations at both the yarn and the fiber levels.  These training
simulations are set up under identical conditions to provide consistent dynamics 
across both levels.


We will elaborate our simulation models and parameter choices in \S\ref{sec:training1}.
In this section, we focus on our main training phase.
In \S\ref{ssec:proc_model}, we briefly revisit the procedural approach for generating cloth models with fiber-level details.
Then, in \S\ref{ssec:learning},  we show how this procedure can be extended to become mechanics-aware by leveraging a custom-design regression neural network.

\subsection{Background: Procedural Modeling of Fiber Geometry}\label{ssec:proc_model}
To generate a cloth model with fiber-level details, a procedural approach originated in textile design has been introduced to graphics recently~\cite{schroder2015image,Zhao:2016:FPY}.
This approach takes as input a set of yarn centerlines and synthesizes fiber curves around each yarn independently.

For a yarn with its centerline following the $z$-axis, each of the constituent fiber~$j$ is modeled as a circular helix:
\begin{equation}
\label{eqn:fiber_curve_0}
\bh_j(z) = \left[r_j\cos\left(\theta_j + \frac{2\pi z}{a}\right),\ 
                 r_j\sin\left(\theta_j + \frac{2\pi z}{a}\right),\
                 z\right]^{\top},
\end{equation}
where $r_j$, $a$ and $\theta_j$ control the radius, pitch and initial phase of the helix $\bh_j$, respectively. 
When building a procedural yarn, the radius $r_j$ of each fiber is drawn independently from a probability distribution
\begin{equation}
\label{eqn:fiber_distrb}
p(r;\ \epsilon, \beta) \propto (1 - 2\epsilon_1)\left( \frac{\E - \E^r}{\E - 1} \right)^{\epsilon_2},
\end{equation}
where $\epsilon_1$ and $\epsilon_2$ are two parameters shared by all fibers.

To better model the migration of fibers, Eq.~\eqref{eqn:fiber_curve_0} can be extended to allow the helix radius $r_j$ to vary with $z$.
A typical implementation of this idea is to have $r_j(z)$ changing under a
sinusoidal pattern between two predefined minimal and maximal values.
Additionally, real-world yarns usually contain multiple substrands, or plies.
In this case, each ply can be described using Eqs.~(\ref{eqn:fiber_curve_0}, \ref{eqn:fiber_distrb}).
Then, the yarn is formed by twisting all the constituent plies around a common centerline.
Please refer to the prior works~\cite{schroder2015image,Zhao:2016:FPY} for more details on fiber migration and multi-ply yarn modeling.

Lastly, given a centerline $\bal$ expressed as a 1D curve parameterized by arc length $z$ with normal $\bn_{\bal}(z)$ and binormal $\bb_{\bal}(z)$, a straight yarn centered at the $z$-axis can be warped to follow this centerline by transforming each fiber $\bh_j$ to $\bh_j^{\bal}$ given by
\begin{equation}
\label{eqn:fiber_curve_1}
\bh_j^{\bal}(z) = \bal(z) + [\bh_j(z)]_x\bn_{\bal}(z) + [\bh_j(z)]_y\bb_{\bal}(z),
\end{equation}
where $[\bh_j(z)]_x$ and $[\bh_j(z)]_y$ denote the $x$- and $y$-components of $\bh_j(z) \in \Real^3$, respectively.

\begin{figure}[t]
	\begin{tabular}{cc}
		\begin{overpic}[width=0.47\columnwidth]{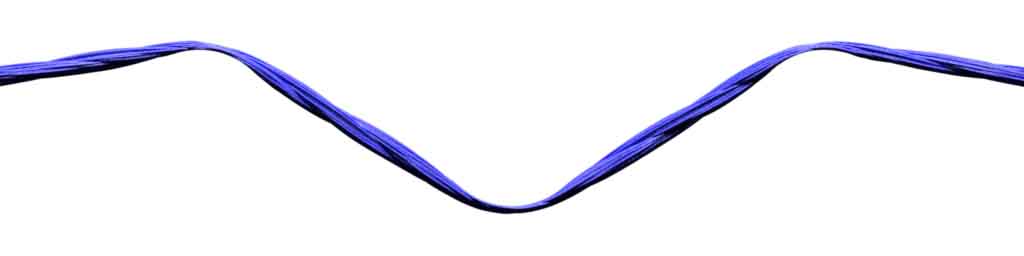}
                    \put(2, 3){\small \bfseries (a) }
		\end{overpic}
		&
		\begin{overpic}[width=0.47\columnwidth]{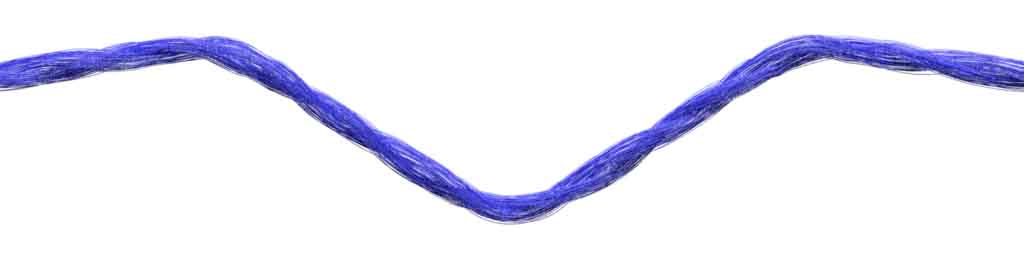}
                    \put(2, 3){\small \bfseries (b) }
		\end{overpic}
	\end{tabular}
    \caption{\label{fig:alter_yarn_dg} 
        \textbf{Deforming fibers using yarn-level deformation gradients} generally leads to unsatisfactory results (b) compared to the reference (a) since they effectively capture average deformation within full grid cells that are much larger than individual fibers.
    }
\end{figure}

\subsection{Learning How Fibers Deform}
\label{ssec:learning}
The procedural model outlined in Eqs.~(\ref{eqn:fiber_curve_0}--\ref{eqn:fiber_curve_1}) does not capture the \emph{deformation} of yarns due to their mechanical responses (e.g., contacting with other yarns and objects or being stretched).

To address this problem, we augment the procedural yarn model with a \emph{cross-sectional affine transformation} $\fiberT \in \Real^{2 \times 2}$ that varies spatially.
Specifically, $\fiberT$ transforms each fiber $\bh_j$ to:
\begin{equation}\label{eqn:cross_sec_comp}
\bh'_j(z) := \begin{bmatrix}
\fiberT(z) & 0\\
0 & 1
\end{bmatrix}
\bh_j(z),
\end{equation}
which can in turn be warped to follow arbitrary yarn centerlines as before using Eq.~\eqref{eqn:fiber_curve_1}.
Based on this formulation, the mechanics-aware modeling of a fabric's fiber microstructures boils down to finding proper cross-sectional transformations $\fiberT$.
This, however, is challenging since cloth fibers usually react to mechanical forces in complicated ways based on their material properties and geometric arrangements.

We aim to obtain the cross-sectional transformations $\fiberT$ using only results simulated at the yarn level.
During these simulations, a yarn is represented as a polyline 
where each segment 
is associated with a deformation gradient $\dg$ that describes the local volumetric deformation.
We provide more details on physics-based simulation of cloth yarns and fibers in \S\ref{sec:training1}.

One na\"ive approach is to directly use the yarn-level deformation gradients to transform the fibers.
Unfortunately, this generally leads to unsatisfactory results, as demonstrated in Figure~\ref{fig:alter_yarn_dg}.

We instead leverage a custom-design regression neural network to infer the cross-sectional transformations $\fiberT$ based on yarn-level deformation gradients $\dg$.
Our goal here is to learn a mapping between the deformation gradients $\dg$ and the cross-sectional deformations $\fiberT$ so that the latter can be used to deform the procedurally generated fibers via Eq.~\eqref{eqn:cross_sec_comp}.
Since different types of yarns (e.g., cotton vs. silk) generally exhibit distinctive fiber structures that cause the yarns to react to external forces differently, we learn the mapping for each yarn separately.

\begin{figure}[t]
	\begin{tabular}{cc}
		\begin{overpic}[width=0.22\textwidth, height=0.05\textheight]{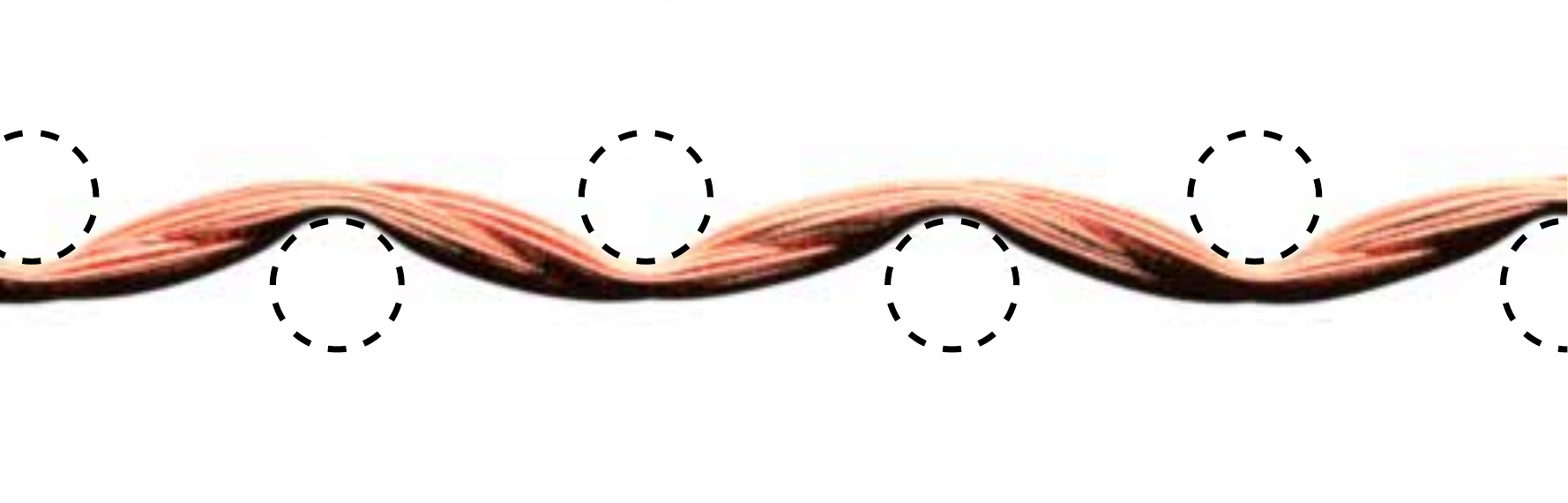}
			\put(2, 3){\color{black} \bfseries pattern 1}
		\end{overpic}
		&
		\begin{overpic}[width=0.22\textwidth, height=0.05\textheight]{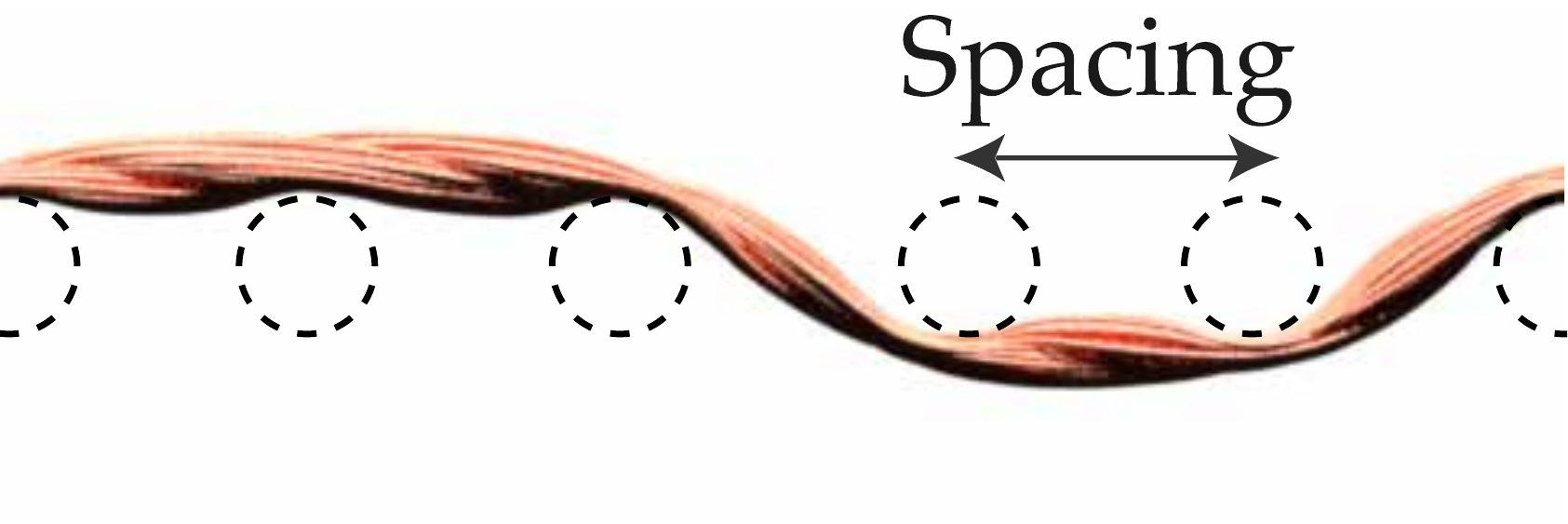}
			\put(2, 3){\color{black} \bfseries pattern 2}
		\end{overpic}
		\\
		\begin{overpic}[width=0.22\textwidth, height=0.05\textheight]{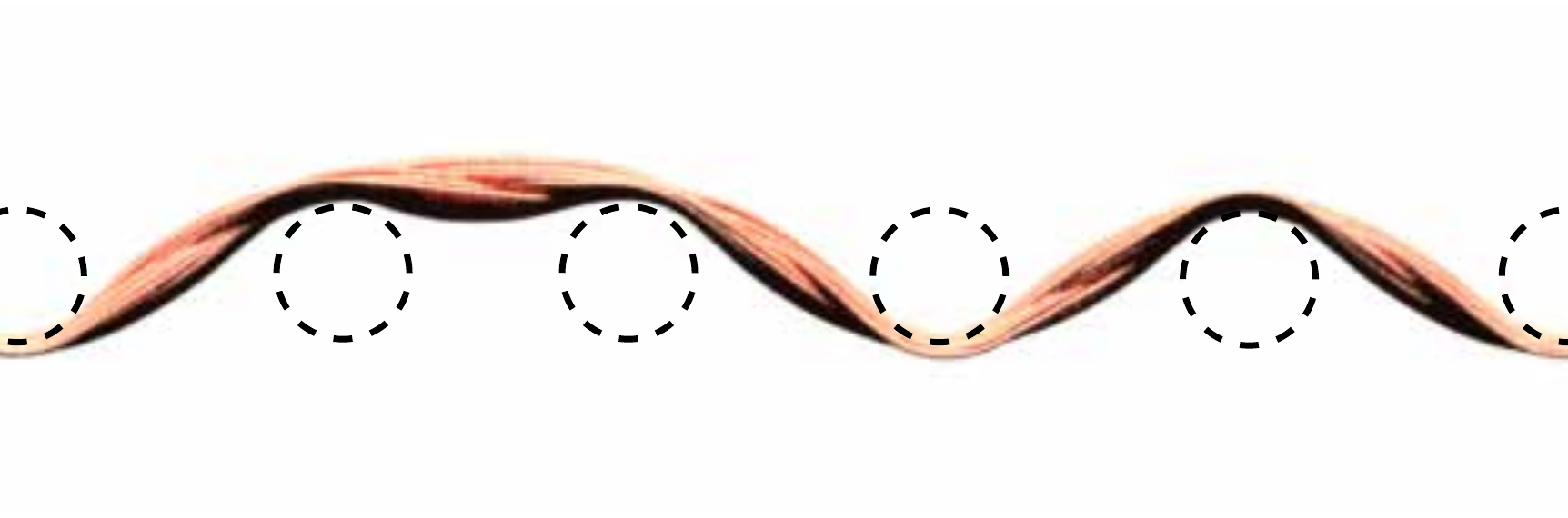}
			\put(1, 3){\color{black} \bfseries pattern 3}
		\end{overpic}
		&
		\begin{overpic}[width=0.22\textwidth, height=0.05\textheight]{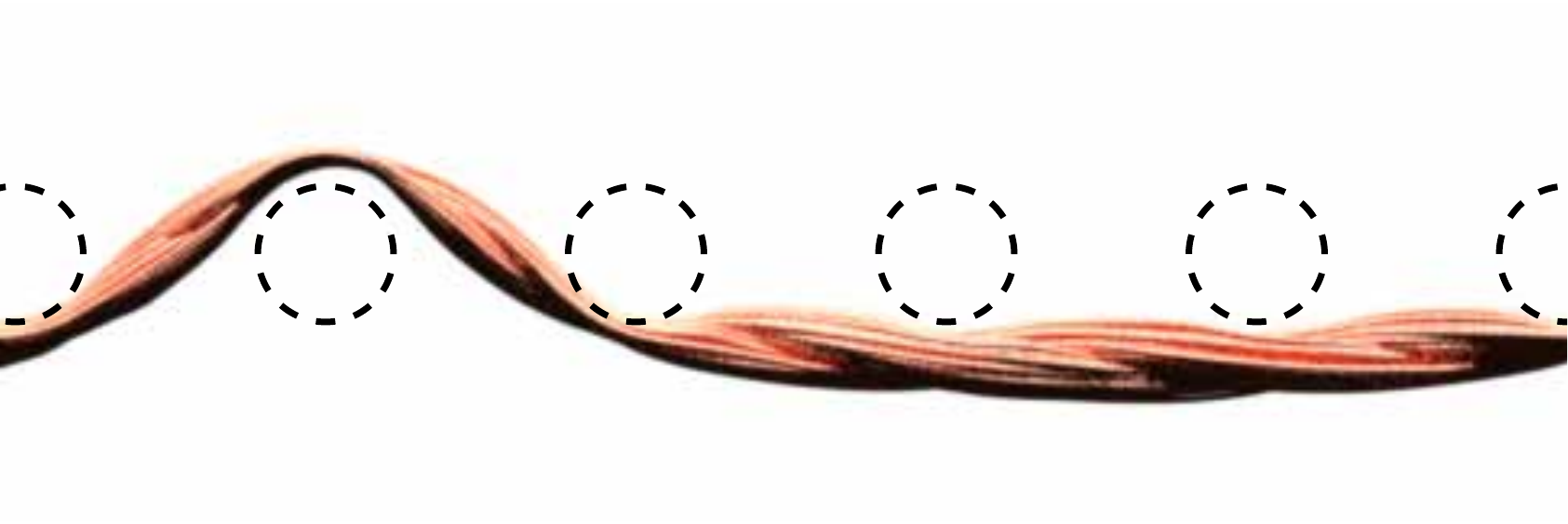}
			\put(2, 3){\color{black} \bfseries pattern 4}
		\end{overpic}
	\end{tabular}
	\caption{\label{fig:training_sims}
		\textbf{Simulation for training.} We simulate the deformation of a yarn under a range of configurations.
		In particular, we use four common patterns~\cite{Zhao:2012:SSP} and six different cylinder spacings per pattern.
		Please refer to the supplemental material for the complete set of configurations.
	}
\end{figure}

\subsubsection{Training data generation}\label{ssec:td}
We generate training data by performing a series of simulations of a single yarn at both the fiber and the yarn levels.
To ensure that the training dataset covers a wide range of yarn and fiber deformations driven by mechanical forces, we configure our training simulations to include a single yarn compressed by a few rigid cylinders.
We use multiple cylinder arrangements for the simulations (Figure~\ref{fig:training_sims}), and collect the results as our training dataset.

\begin{figure}[t]
	\centering
	\includegraphics[width=0.99\columnwidth]{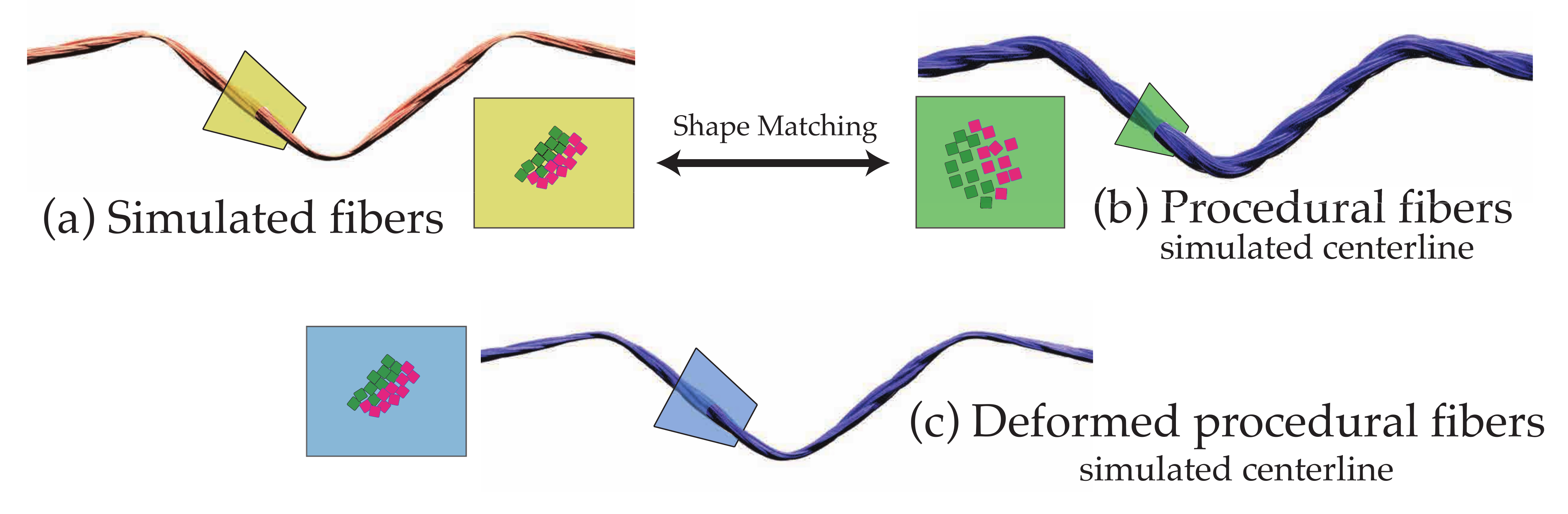}
	\caption{\label{fig:dual_trainings}
	\textbf{Computing cross-sectional fiber deformations.}
		We simulate a yarn at both fiber~(a) and yarn~(b) levels.  At the yarn level, we procedurally generate fibers following the simulated centerline (which agrees with the fiber-level simulation thanks to the parameter homogenization described in \S\protect\ref{sec:training1}).
		For each cross section along both yarns (from yarn- and fiber-level simulations), we compute the intersections
		of the fibers on the cross-sectional plane to obtain a 2D point cloud per yarn.
		Lastly, we perform 2D shape matching between the two point clouds, and obtain the desired fiber deformation matrix $\fiberT$.
		By deforming the procedurally generated fibers, the resulting fiber microstructure~(c) matches the results of fiber-level simulation closely.
	}
\end{figure}

\paragraph*{\em Training simulations}
For each simulation configuration, as shown in Figure~\ref{fig:dual_trainings}-a, we use the procedural approach depicted in
\S\ref{ssec:proc_model} to generate the constituent fibers of a straight yarn as the initial state of the simulation.
Then, we quasi-statically move a few rigid cylinders toward the yarn and simulate the mechanical response of the fibers.  
We record the shape of all fiber curves after each time step.

At the yarn level, we simulate how a yarn centerline reacts to the same quasi-statically moving cylinders (Figure~\ref{fig:dual_trainings}-b).
Thanks to our parameter homogenization step (which will be detailed in \S\ref{sec:training1}), the simulated yarn centerline closely agrees with that from the fiber-level simulation.
After each time step, we record the yarn centerline as well as the normal and binormal directions (computed using the simulated twisting angles) and the deformation gradients along the yarn.
Using the recorded centerline and normals, we generate fiber curves following the simulated centerline using
Eqs.~(\ref{eqn:fiber_curve_0}--\ref{eqn:fiber_curve_1}) with no deformation. 

\paragraph*{\em Shape matching}
Given results from the simulations at both levels, we now compute $\bm{T}(z)$ along the yarn centerline to compensate for the discrepancy between the simulated fiber curves and the procedurally generated ones.
We discretize $\bm{T}(z)$ at the edge centers of the yarn polyline. 
At each edge center, we define a cross-sectional plane perpendicular to the tangential direction along the centerline.
Then, we compute the intersections between the cross-sectional plane and the fiber curves~(\figref{dual_trainings}-ab).
Let $\mathcal{P}_{\text{fiber}}$ and $\mathcal{P}_{\text{yarn}}$ denote the cross-sectional fiber locations obtained using simulated fiber curves and procedurally generated ones, respectively.
Then, the desired cross-sectional transformation $\fiberT$ should warp $\mathcal{P}_{\text{yarn}}$ to match $\mathcal{P}_{\text{fiber}}$ as closely as possible.
This can be formulated as a shape matching problem~\cite{Muller2005MDB} in 2D. 
By applying the computed deformations $\fiberT$ via Eq.~\eqref{eqn:cross_sec_comp}, the procedurally
generated fibers can closely match the fully simulated fibers (\figref{dual_trainings}-c). 

\subsubsection{Our regression network}\label{sssec:our_NN}
Next, we train a regression neural network to learn the relation between deformation gradients simulated at the yarn level and cross-sectional transformations of cross-sectional fiber locations.
At runtime, the latter will be estimated using the network \emph{without} resort to fiber-level simulations.  

\begin{figure}[t]
	\centering
	\includegraphics[width=0.6\columnwidth]{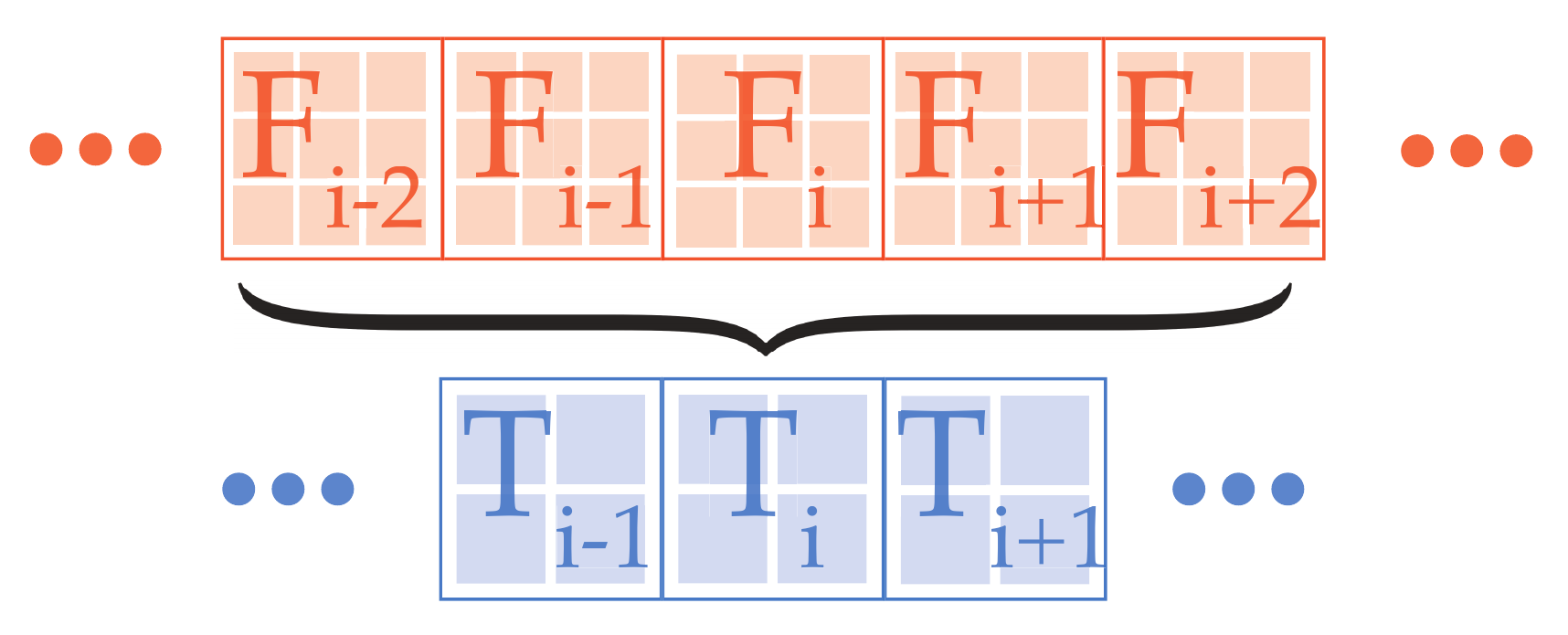}
	\caption{\label{fig:NN_io}
		\textbf{Input and output} to our neural network.
		Our neural network takes as input deformation gradients given by yarn-level simulation
		in a small neighborhood window around a cross section, and estimates the
		cross-sectional deformation at that cross section.
	}
\end{figure}

\paragraph*{\em Input and output}
The deformation $\fiberT$ at any cross section along the yarn mostly depends on the external forces acting at a small neighborhood around the cross section.
Thus, when our regression network estimates $\fiberT$ at a cross section, it takes as input the simulated deformation gradients in a small window around that cross section (Figure~\ref{fig:NN_io}).
In other words, our network acts like a (nonlinear) 1D filter along yarn centerlines.

\paragraph*{\em Transforming training data to local frames}
Since simulated yarn curves can be rotated without affecting their internal structures, it is desired for the input and output of our regression network to be rotationally invariant.  
Yet, our deformation gradients do vary under rotations since they effectively transform individual segments of the yarn polyline from their canonical material spaces (where the tangential direction is aligned with the $x$-axis) to the world space.
This observation suggests that we should transform the simulated deformation gradients into a \emph{local} frame of reference to ensure rotational invariance.

Consider a local neighborhood of cross sections (\figref{NN_io}), we define the local frame of reference by specifying the tangent $\bt$, normal $\bn$, and binormal $\bb$ at each cross section in the neighborhood.
The tangent $\bt$ naturally follows that of the yarn centerline.
To determine the normal $\bn$, we use the principal normal $\bn_0$ at the central cross section. 
Let $\bt_0$ be the tangent at this location and $\bb_0 := \bn_0 \times \bt_0$. 
Then, $(\bt_0, \bn_0, \bb_0)$ defines the local frame of reference%
\footnote{Note not to confuse these frames with ones used for procedural fiber generation~\eqref{eqn:fiber_curve_1}: unlike the latter ones that typically vary smoothly along a yarn centerline, the former frames depend on (and only on) the local yarn geometry around individual cross sections and can differ considerably even between overlapping neighborhoods (e.g., when the yarn curvature changes sign).}
at the central cross section.
Using this frame as a reference, we then determine the local frames for the rest of cross sections in the neighborhood as follows.
For a cross section with a tangent direction $\bt'$, we set its normal and binormal as
\begin{equation}
\label{eqn:norm_binorm}
\bn' := \bm{S}(\bt_0 \to \bt') \,\bn_0 \;\text{and}\;
\bb' := \bm{S}(\bt_0 \to \bt') \,\bb_0, 
\end{equation}
where $\bm{S}(\bt_0 \to \bt') \in \Real^{3 \times 3}$ denotes the minimal rotation that aligns $\bt_0$ with $\bt'$.
As a corner case, when the principal normal at the central cross section vanishes, we find a nearest neighbor with non-vanishing principal normal and generate the reference frame $(\bt_0, \bn_0, \bb_0)$ there.
As we define the normal directions in the rotation-minimizing fashion, the local frame of reference changes smoothly along the centerline within any neighborhood.

Next, we transform the deformation gradients into the local frames of reference using
\begin{equation}
\label{eqn:dg_local}
\dglocal = [\bt,\ \bn,\ \bb] \,\dg,
\end{equation}
where $\dg$ is the original deformation gradient provided by the yarn-level simulation.
Then, all the $\dglocal$ matrices together in a neighborhood serve as an input to the regression network.

Lastly, since the normal and binormal directions given by Eq.~\eqref{eqn:norm_binorm} generally differ from those given by the yarn centerline, we need to also transform our shape matching result $\fiberT$ at each cross section into our newly defined local spaces. 
Let $\bm{S}(\bn_{\alpha} \to \bn) \in \Real^{2 \times 2}$ denote the rotation that transforms $\bn_{\alpha}$ to $\bn$.
Then, the cross-sectional deformation local to the neighborhood is
\begin{equation}
\label{eqn:shape_match_local}
\fiberTlocal = \bm{S}(\bn_{\alpha} \to \bn) \, \fiberT \, \bm{S}(\bn_{\alpha} \to \bn)^{\top},
\end{equation}
which is also the expected output from our regression network.

\begin{figure}[t]
	\centering
	\includegraphics[width=0.99\columnwidth]{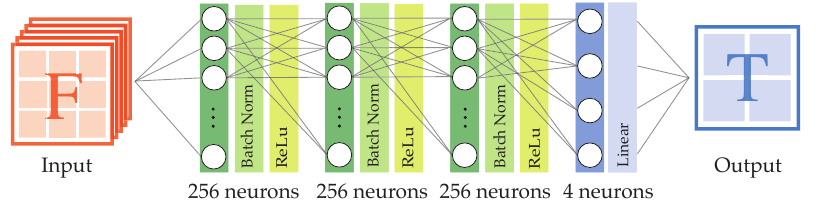}
	\caption{\label{fig:NN_architecture}
		\textbf{Network architecture.}
		Our regression neural network involves three fully connected hidden layers with 256 nodes each (shown in green).
		The network acts as a 1D nonlinear filter along a yarn centerline and turns simulated deformation gradients~$\dg$ into cross-sectional fiber deformations~$\fiberT$ (see Figure~\protect\ref{fig:NN_io}).
	}
\end{figure}

\paragraph*{\em Loss function for training}
After obtaining the training data involving local deformation gradients $\dglocal \in \Real^{3 \times 3}$ and fiber deformation $\fiberTlocal \in \Real^{2 \times 2}$, we train our regression network by minimizing the following loss function:
\begin{equation}
\small
\label{eqn:loss}
\underbrace{\left\| \regnet\left(\{ \dglocal \}_i \right) - \fiberTlocal_i \right\|_2^2}_{\text{parameter loss}} +
\lambda \underbrace{\left\| \regnet\left(\{ \dglocal \}_i \right) \bm{H} - \fiberTlocal_i \, \bm{H} \right\|_2^2}_{\text{regularization}},
\end{equation}
where $\{ \dglocal \}_i$ denotes the local deformation gradients from the neighborhood around the $i$-th cross section, $\fiberTlocal$ is the desired fiber deformation at this cross section, $\regnet()$ indicates the network output (given the deformation gradients as input), $\bm{H}$ is a constant matrix containing the 2D location of each fiber 
center averaged over all cross sections, and $\| \cdot \|_2$ denotes the 2-norm of a matrix.
Intuitively, this loss function involves a parameter loss term that captures the difference between network prediction and the desired output.
Additionally, to regularize the network, we introduce an extra regularization term weighted by some $\lambda > 0$.
As the $\bm{H}$ matrix is comprised of the cross-sectional locations of a set of fibers, $\regnet\left(\{ \dglocal \}_i \right) \bm{H}$ and $\fiberTlocal_i \, \bm{H}$ give the transformed fiber locations using the network prediction and the desired output, respectively.
The regularization term then captures the difference between these two transformed 2D point clouds.

\paragraph*{\em Data augmentation}
To further improve the robustness of our regression network, we augment our training data by a factor of four.
Given a set of local frames $\{(\bt_k, \bn_k, \bb_k)\}$ in a neighborhood (where $k$ indices each cross section in the neighborhood), we compute three additional sets of frames of reference given by $(\bt_k, -\bn_k, -\bb_k)$,
$(-\bt_k, \bn_k, -\bb_k)$, and $(-\bt_k, -\bn_k, \bb_k)$ for each $k$.
For each set of frames, we then compute the corresponding local deformation gradients $\dglocal$ and fiber transformations $\fiberTlocal$ via Eqs.~(\ref{eqn:dg_local}, \ref{eqn:shape_match_local}) and include them in the training dataset.
In practice, we found that this data augmentation step greatly improves the stability of our regression network.

\paragraph*{\em Network implementation}
Our regression network consists of three fully connected hidden layers with 256 internal nodes each (see Figure~\ref{fig:NN_architecture}).
We implement this network using the Python-based Keras library~\cite{keras} with Tensorflow~\cite{tensorflow}.
We split our augmented data into a training set (85\%) and a validation set (15\%), and optimize the network weights using the Adam method~\cite{Kingma2014AdamAM}.

\section{Simulation Model}\label{sec:training1} 
This section presents our numerical model for both yarn- and fiber-level simulation.
The simulation model serves two purposes, i) to 
generate training data for the deep neural network enriching fiber-level geometries (\S\ref{ssec:learning}),
and ii) to perform at runtime yarn-level simulation providing input to the trained 
deep neural network (\secref{runtime}).

To this end, not only does the simulation model need to capture both yarn- and
fiber-level dynamics, the simulated yarn dynamics must also be consistent across both scales.
This poses a problem of fitting (or homogenizing) yarn-level material parameters from the fiber-level simulation. 
While several models have been proposed to simulate yarns, none of them addresses the problem 
of parameter homogenization---one that this section focuses on.

\subsection{Unified Simulation at Yarn and Fiber Level}\label{sec:sim}
We simulate textile at both yarn and fiber level in a unified framework,
which uses the elastic rod model~\cite{bergou2008discrete} and a
hybrid Lagrangian/Eulerian method based on~\cite{jiang2017anisotropic}.
We refer to their papers for more details, while briefly reviewing the key components
here for describing our parameter homogenization. 

\begin{figure}[t]
	\centering
	\includegraphics[width=0.81\columnwidth]{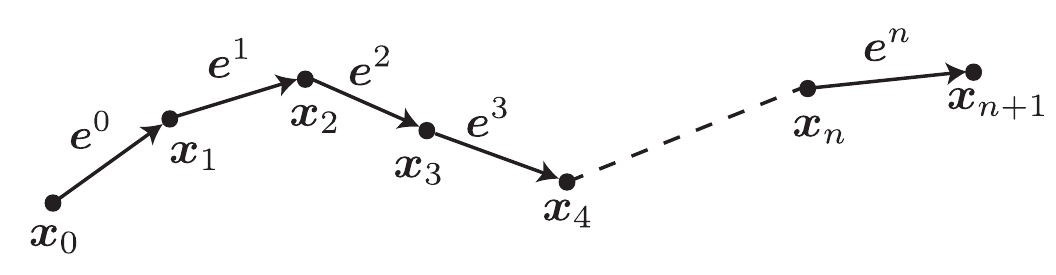}
	\caption{\label{fig:polyline}
		\textbf{Yarn/fiber discretization.}
		In yarn-level simulation, each yarn is represented by a series of particles connected on a polyline. In fiber-level simulation, each fiber is represented as a polyline, and a group of fibers are twisted to form a yarn.
	}
\end{figure}

\paragraph*{\em Internal force}
In fiber-level simulation, each fiber is represented as a polyline; 
and in yarn-level simulation, a polyline represents a yarn. In both cases, we compute 
the internal force using the elastic rod model~\cite{bergou2008discrete}.
The internal force is contributed by three types of elastic energies, namely, the stretching, bending, and twisting energy.
Following the notation in~\cite{bergou2008discrete} (summarized in \figref{polyline}),
the discrete stretching is defined as 
\begin{equation}\label{eq:stretch}
    E_{\text{stretch}} = \sum_{i=0}^n\frac{k}{2}\left(|\bm{e}^i|-|\bar{\bm{e}}^i|\right)^2,
\end{equation}
where $|\bm{e}^i|$ is the (stretched) edge length while $|\bar{\bm{e}}^i|$ is
the corresponding rest length, and $k$ is the stretching coefficient. 
In addition, the discrete bending and twisting energy are given by
\begin{equation}\label{eq:bend}
    E_{\text{bend}} = \sum_{i=1}^n\frac{\alpha(\bm{\kappa b}_i)^2}{2\bar{l}_i}
    \;\;\textrm{and}\;\;
    E_{\text{twist}} = \sum_{i=1}^n\frac{\beta m_i^2}{2\bar{l}_i}.
\end{equation}
Here, $\bar{l}_i$ is the effective rest length of particle $i$, defined as 
$\bar{l}_i = \left(|\bar{\bm{e}}^{i-1}|+|\bar{\bm{e}}^i|\right)/2$.
The scalar $m_i$ is the twisted angle of the polyline at particle $i$,
and $\bm{\kappa b}_i$ is its curvature binormal. 
We refer to~\cite{bergou2008discrete} for their specific formulas. 
Lastly, $\alpha$ and $\beta$ are the bending and twisting coefficients, respectively.
The parameters $\alpha$, $\beta$, as well as $k$ (in Eq.~\eq{stretch}) are material
related. They are different in fiber- and yarn-level simulations:
the values in yarn-level simulation are the ``homogenized'' version of those in
fiber-level simulation.

\paragraph*{\em External force}
Fibers (or yarns) are deformed by external forces such as collision and friction forces.
Yet, resolving collision and friction is computationally prohibitive,
especially in fiber-level simulation, wherein
yarns are modeled explicitly as a large number of fibers.
We therefore take a hybrid Lagrangian/Eulerian approach~\cite{jiang2017anisotropic},
which treats the deformation of fibers (or yarns) volumetrically.
In this approach, collision forces are modeled as resistance of 
compression of the fiber (or yarn) volume, 
and friction force is viewed as the resistance of shearing of the fiber (or yarn) volume.
To model these forces in the discretized setting,
each edge $\bm{e}^i$ is associated with an elastic deformation gradient
$\bm{F}^i$, which describes the local deformation of the fiber (or yarn) volume
near $\bm{e}^i$. Particularly, the local material directions
$\bm{D}^i =\left[\bm{D}_1\;\bm{D}_2\;\bm{D}_3\right]$ of $\bm{e}^i$ at the rest state
is deformed into $\bm{F}^i\bm{D}^i$, the deformed material directions.
Applying the QR-decomposition, $\bm{Q}\bm{R} = \bm{F}^i\bm{D}^i$, 
yields the orthonormal material directions $\bm{Q}$ and the matrix $\bm{R}$.
The latter can be further decomposed into three components (i.e.,
$\bm{R}=\bm{R}_3\bm{R}_2\bm{R}_1$) given as follows:

\begin{equation}\label{eq:RRR}
\small
\bm{R}_1 = 
\begin{bmatrix}
    r_{11} & 0 & 0      \\
    0 	   & 1 & 0      \\
    0 	   & 0 & 1      \\
\end{bmatrix}
\,
\bm{R}_2 = 
\begin{bmatrix}
    1      & r_{12} & r_{13}      \\
    0 	   & 1 & 0      \\
    0 	   & 0 & 1 		\\
\end{bmatrix}
\,
\bm{R}_3 = 
\begin{bmatrix}
    1      & 0 & 0      \\
    0 	   & r_{22} & r_{23}      \\
    0 	   & 0 & r_{33} \\
\end{bmatrix}.
\end{equation}
$\bm{R}_1$ describes the stretch along the tangential direction of the curve (be it a fiber 
or yarn), $\bm{R}_2$ describes the shear along the curve, and 
$\bm{R}_3$ indicates the cross-sectional deformation.  
Similar to~\cite{jiang2017anisotropic}, we use two quadratic energy functions
$g(\bm{R}_2)$ and $h(\bm{R}_3)$ to model the energy generated by collision and friction
forces. Their specific forms will be described shortly in \secref{param}.
Different from~\cite{jiang2017anisotropic}, we neglect the stretching component $\bm{R}_1$, as 
the stretching energy is already taken into account in the aforementioned elastic rod model,
and the method of~\cite{jiang2017anisotropic} does not consider the bending and twisting 
forces of the curves.

With these energy definitions, the internal and external forces can be computed by taking
the derivatives of the energies with respect to particle positions of the polylines.
Complete derivations of these forces are detailed in~\cite{bergou2008discrete} and~\cite{jiang2017anisotropic}.
In our simulation, we use the Explicit Symplectic Euler (with a time step of $2 \times 10^{-5}$) to advance the states 
of the curves over time. 

\paragraph*{\em Remark}
The volumetric view of the external forces offers another advantage for us. The deformation gradient $\bm{F}^i$
at each edge $\bm{e}^i$ encapsulates the local deformation (such as stretch, rotation, and shear) due to external forces.
Thereby, in yarn-level simulation, $\bm{F}^i$ provides a concise clue about how the fibers that form the yarn should deform.
Naturally, $\bm{F}^i$ will be fed into the deep neural network for estimating fiber-level geometries, as has been discussed in \S\ref{ssec:learning}.

\begin{figure}[t]
	\centering
	\includegraphics[width=0.85\columnwidth]{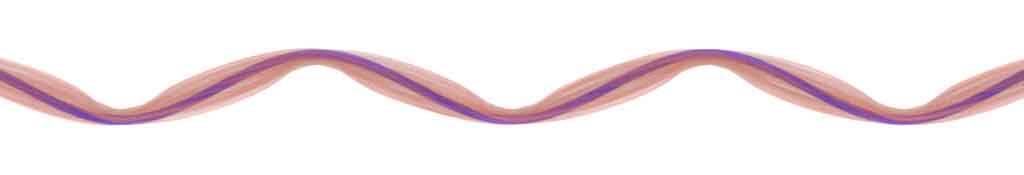}
	\caption{\label{fig:centerline}
		\textbf{Centerline of a fiber bundle.} 
		A yarn is simulated at fiber level with individual fibers shown in blue.
		The yarn's centerline is highlighted in red. We fit yarn-level parameters so that the
		yarn curve resulting from yarn-level simulation closely matches the centerline from fiber-level simulation.
	}
\end{figure}

\subsection{Parameter Homogenization}\label{sec:param}
We use the above force models in both fiber- and yarn-level simulations.  To ensure
consistent yarn dynamics across the two levels, we must set proper material
parameters in Eqs.~\eq{stretch} and~\eq{bend} and the energy functions, $g(\bm{R}_2)$ and
$h(\bm{R}_3)$. In fiber-level simulation, material parameters are set
based on the physical properties 
of the fibers (such as silk and cotton). 
In yarn-level simulation,
we need to set the parameters so that the simulated yarn curves, under the same initial condition, 
closely match the centerlines of the corresponding fiber bundles in the fiber-level simulation (\figref{centerline}).

%

\subsubsection{Fitting parameters of internal forces}
\begin{figure}[t]
	\centering
	\includegraphics[width=0.99\columnwidth]{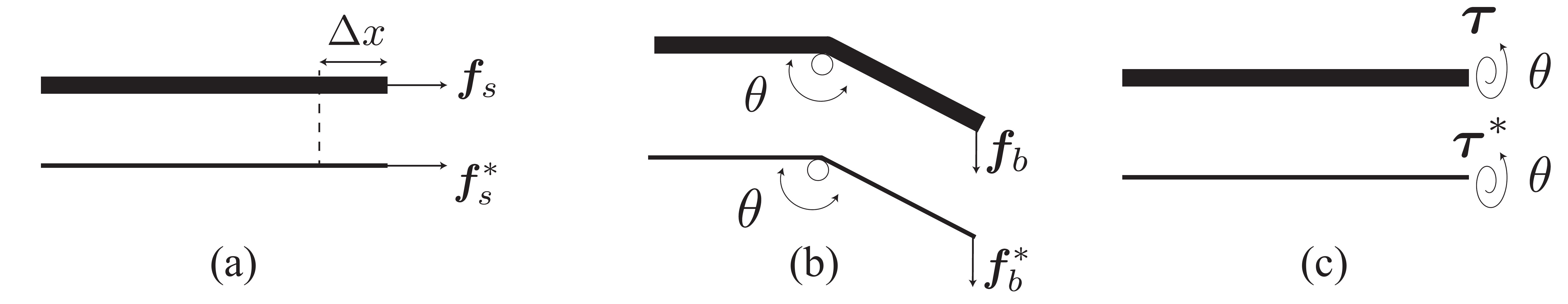}
	\caption{\label{fig:fitting}
		\textbf{Fitting elastic rod parameters.}
		\textbf{(a)} Fixing one end of a yarn, we stretch it by $\Delta x$ in both fiber-level (top) and yarn-level (bottom) simulation.
		The stretching coefficient $k$ is found in a binary search to match the resulting stretching forces $\bm{f}^*_{\text{s}}$ and $\bm{f}_{\text{s}}$.
		\textbf{(b)} Fixing one end and the middle point of a yarn, we quasistatically move the other end to bend the yarn 	in both fiber-level (top) and yarn-level (bottom) simulations by the same amount.
		The bending coefficient $\alpha$ is found in a binary search to match the resulting bending forces $\bm{f}^*_{\text{b}}$ and $\bm{f}_{\text{b}}$ at the end.
		\textbf{(c)} Similarly, we fix one end of the yarn, and twist at the other end by an angle $\theta$
		in fiber-level (top) and yarn-level (bottom) simulations. The twisting coefficient $\beta$ is obtained in a binary search to match the resulting torques $\bm{\tau}^*$ and $\bm{\tau}$.
	}
\end{figure}

We first decide elastic rod parameters ($k$, $\alpha$, and $\beta$) in Eqs.~\eq{stretch} and~\eq{bend} in yarn-level simulation.
Because these parameters are independent from each other, we fit their values separately and in a similar fashion.
To fit $k$, we quasistatically stretch a fiber bundle by a small distance $\Delta x$ 
in the fiber-level simulation,
and evaluate the collective stretching force $\bm{f}^*_{\text{s}}$ at the end of the fiber bundle.
Meanwhile, we quasistatically stretch a yarn of the same length by $\Delta x$ in the yarn-level simulation.
We search for $k$ such that the stretching force $\bm{f}_{\text{s}}$ at the end the yarn matches $\bm{f}^*_{\text{s}}$. 
Because $\bm{f}_{\text{s}}$ is proportional to $k$, the value of $k$ can be quickly
found through a binary search (\figref{fitting}-a). 
In practice, we find that $k$ is not sensitive to the choice of $\Delta x$,
and we repeat this fitting process multiple times each with a different $\Delta x$ value, and then take the average.
The bending and twisting coefficients ($\alpha$ and $\beta$) are obtained in a similar way, as shown in \figref{fitting}-b and -c.

%
%

\subsubsection{Fitting parameters of external forces}
As introduced in \secref{sim}, the external forces are determined by two energy functions $g(\bm{R}_2)$ and $h(\bm{R}_3)$
both having quadratic forms. In fiber-level simulation, we use
$g(\bm{R}_2) = \gamma(r_{12}^2+r_{13}^2)$ and $h(\bm{R}_3)=\mu(r_{22}^2+r_{33}^2)+\frac{\lambda}{2}(r_{22}+r_{33})^2$
similar to~\cite{jiang2017anisotropic}. Here, $r_{ij}$ are matrix elements defined in Eq.~\eq{RRR}.
$\gamma$, $\mu$, and $\lambda$ are parameters set according to specific type of fibers.
In yarn-level simulation, we use 
\begin{equation}\label{eq:qform}
    \begin{split}
g(\bm{R}_2) &= a_1 r_{12}^2+a_2 r_{12}r_{13} + a_3 r_{13}^2 \;\textrm{and}\\
h(\bm{R}_3) &= b_1 r_{22}^2+b_2 r_{22}r_{33} + b_3 r_{33}^2,
    \end{split}
\end{equation}
where parameters $a_i$ and $b_i$ ($i=1,2,3$) are what we need to fit.

Fitting these yarn-level parameters is nontrivial.  Unlike the elastic rod
parameters, the parameters $a_i$ and $b_i$ are related to the external forces
through nonlinear derivatives of $g(\bm{R}_2)$ and $h(\bm{R}_3)$; their
relationships to the forces can not be decoupled, and neither are they
proportional to the forces. Consequently, the binary search is not useful. 
When choosing $a_i$ and $b_i$, we need to ensure the energy 
functions $g(\bm{R}_2)$ and $h(\bm{R}_3)$ are positive semidefinite---a constraint
known as the \emph{conic constraint} in convex optimization. 
Moreover, in fiber-level simulation, the deformation gradients (and
hence $\bm{R}_2$ and $\bm{R}_3$) is defined on each finite elements $\bm{e}^i$
of the fibers, whereas in yarn-level simulation they are defined on the
elements of yarns. 
Caution is needed in estimating yarn-level deformation gradients 
from fiber-level simulation.

Our parameter homogenization is based on a key observation. 
Consider a yarn resulted from a fiber-level simulation. 
Its centerline is what the yarn-level simulation should produce when material
parameters are fit properly. Since the elastic rod parameters of yarn-level simulation have already been
determined in the previous step, the expected internal forces in yarn-level simulation are known.
In general, knowing the internal forces is insufficient to infer the yarn-level external forces.
However, if the yarn is in \emph{static equilibrium}, internal forces must be balanced by external forces.
This means that the expected external forces are known as well.

In light of this observation, we use fiber-level simulation to obtain a yarn's static equilibrium state under
collisions, and thereby compute the external forces needed in yarn-level simulation.
We then estimate yarn-level deformation gradients, which are in turn used to 
fit the parameters $a_i$ and $b_i$ ($i=1,2,3$) in Eq.~\eq{qform}.

\begin{figure}[t]
	\centering
	\includegraphics[width=0.99\columnwidth]{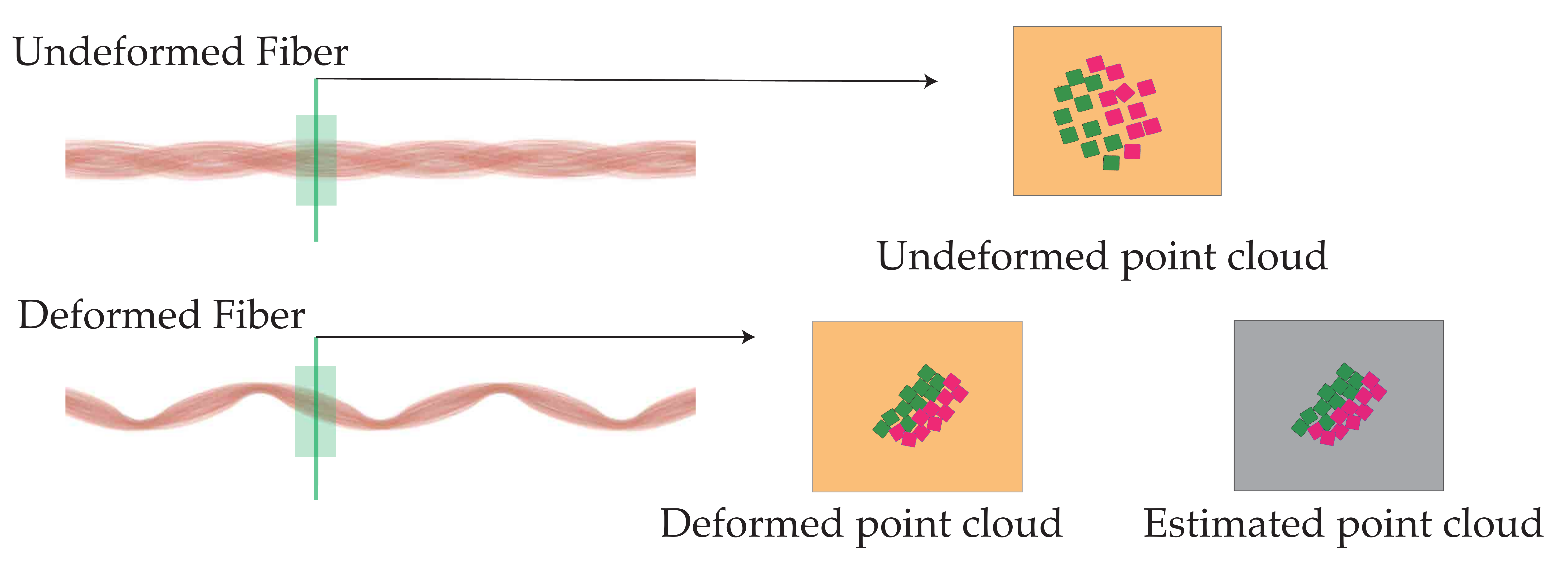}
	\caption{\label{fig:estimating}
		\textbf{Estimating deformation gradient.} 
		Consider a small segment of a yarn (highlighted in red box).
		We visualize the cross-sectional fiber distributions on the right.
		In fiber-level simulation, the fiber distribution can get severely deformed (bottom-right).
		We estimate a ``homogenized'' deformation gradient (i.e., a $3\times3$ transformation matrix) that
		transform the fibers in the rest material space to match the deformed fibers as closely as possible.
	}
\end{figure}

\paragraph*{\em Estimating deformation gradients}
We now describe how to estimate yarn-level deformation gradients from fiber-level simulation results.
As illustrated in \figref{estimating}, consider a small segment of a yarn. In fiber-level simulation, this segment 
consists of a set of fiber particles distributed over a group of fibers. Let $\{\bar{\bm{x}}_j\}_{j=1}^n$
denote the positions of these particles at the rest state, and $\{\bm{x}_j\}_{j=1}^n$ be their corresponding position 
at the deformed state. The yarn-level deformation gradient $\bm{F}$ at the same segment should be able to transform 
$\{\bar{\bm{x}}_j\}_{j=1}^n$ into $\{\bm{x}_j\}_{j=1}^n$. This is again a shape matching problem~\cite{Muller2005MDB,Zheng2013OEM} 
similar to that we encountered in \S\ref{ssec:td}.
$\bm{F}$ is obtained by solving a small least-squares problem,
\begin{equation}\label{eq:sm}
    \bm{F} = \arg\min_{\bm{F},\bm{t}} \sum_{j=1}^n \|\bm{F}(\bm{x}_j - \bm{t}) - \bar{\bm{x}}_j\|_2^2,
\end{equation}
where $\bm{t}$ is a vector to eliminate the translational difference, jointly optimized with $\bm{F}$.

\paragraph*{\em Fitting $a_i$ and $b_i$}
External forces are related to $a_i$ and $b_i$ through the derivatives of $g(\bm{R}_2)$ and
$h(\bm{R}_3)$ with respect to the deformation gradient $\bm{F}$, because $\bm{R}_2$ and $\bm{R}_3$ are computed using $\bm{F}$.
Once the yarn-level external forces and deformation gradients are estimated, we are able to fit the parameters $a_i$ and $b_i$. 
A complete exposition of the fitting process involves detailed derivations from the Lagrangian/Eulerian approach~\cite{jiang2017anisotropic}.
Without interrupting the presentation flow, we defer it in the appendix, 
a six-dimensional semidefinite-quadratic-linear programming problem.

\section{Run-Time Phase}\label{sec:runtime}

With the preprocessing stages completed, we are now able to generate animated
cloth models with fiber-level details.  Our runtime pipeline takes the
following inputs:
\begin{itemize}
\item The yarn-level material parameters obtained in the parameter fitting step (\S\ref{sec:training1});
\item The parameters for procedural generation of fibers via Eqs.~\eqref{eqn:fiber_curve_0} and \eqref{eqn:fiber_distrb};
\item The pre-trained neural network that maps yarn-level deformation gradients to fiber-level deformations along yarn centerlines (\S\ref{sec:training2}).
\item The initial condition of a fabric for performing yarn-level simulation. 
\end{itemize}
The output of our system at runtime is a collection of animated fiber curves.

Our yarn-level simulation outputs the shape and arrangement of yarn centerlines together with
the deformation gradients and normal directions along the yarn.  
With these information, we first procedurally generate fiber curves using the simulated yarn centerline and
normals without deformation.  

As demonstrated in Figure~\ref{fig:motivating_example}, neglecting
fiber mechanics leads to inaccurate macroscopic appearance.
Therefore, we update the cross-sectional fiber locations using
deformation matrices $\bm{T}$ estimated by our neural network.
Concretely, we apply our neural network as a 1D filter along every yarn of the fabric. For
each cross section window, we first transform the simulated deformation gradients into
corresponding local frames of reference (\S\ref{sssec:our_NN}), and feed them to the neural network to estimate
the cross-sectional deformation matrix $\bT_{\text{local}}$.
Via the inverse of Eq.~\eqref{eqn:shape_match_local}, this matrix $\bT_{\text{local}}$ is transformed back to $\bT$, which is in turn 
used to transform the fiber curves in Eq.~\eqref{eqn:cross_sec_comp}.
Finally, the fiber-level geometries are ready for rendering the fabric.

\section{Results}
\label{sec:results}
\newlength{\resLen}
\begingroup
\setlength{\columnsep}{10pt}%
\begin{wraptable}[11]{R}{3.5cm}
	\vspace{-2mm}
	\centering
	\caption{Material parameters of fiber-level simulation.}\label{tab:param}
	\vspace{-2mm}
	\begin{tabular}{cc} 
	   \whline{0.8pt}
	 Param. & Value  \\  \hline
	   \whline{0.05pt}
	 $k$ & $1.87 \times 10^{11}$  \\ 
	 $\alpha$ & $4.07 \times 10^5$  \\ 
	 $\beta$ & $1.08 \times 10^3$  \\ 
	 $\gamma$ & $9.60 \times 10^5 $\\ 
	 $\mu$ & $ 3.51 \times 10^5$\\
	 $\lambda$ & $1.01 \times 10^6$ \\
	   \whline{0.8pt}
	\end{tabular}
	\vspace{3mm}
\end{wraptable}
To generate the training dataset, we simulate a single yarn at both fiber and yarn levels.
The material parameters used in fiber-level simulation is listed in Table~\ref{tab:param}.
While we use the same set of fiber-level parameters, the fitted yarn-level parameter set may differ, because the yarn-level parameters
also depend on the specific fiber structure in the yarn.
Nevertheless, the parameters in yarn-level simulation are automatically decided using our parameter fitting algorithm (\secref{param}).
At runtime, only yarn-level simulation is needed.
In principle, our technique is independent of the underlying yarn-based simulation method.
In practice, we use the anisotropic elastoplasticity model introduced by Jiang~et~al.~\cite{jiang2017anisotropic} for its high physical accuracy. 
Our training fiber- and yarn-level simulations take 8 days to run on a single workstation (for all configurations).
Upon obtaining the training dataset, our regression neural network is trained in 30 minutes.

To render our mechanics-aware cloth models, we fully realize them by procedurally generating all the fibers and deforming them using cross-sectional transformations predicted by our regression network.
Our method could be combined with previous realization-minimizing techniques~\cite{Luan:2017:FOP} for better rendering performance.
\endgroup

\subsection{Validation and Evaluation}
\label{ssec:res_eval}
\begin{figure*}[!t]
	\centering
	\small
	\addtolength{\tabcolsep}{-4pt}
	\setlength{\resLen}{0.24\textwidth}
	\begin{tabular}{ccccc}
		& 
		(a1) Photograph
		&
		(b1) Fiber-level sim.
	    &
	    (c1) Ours
	    &
	    (d1) Baseline
	    \\
	    \multirow{2}{*}[18pt]{\rotatebox[origin=c]{90}{Polyester (2-ply)}}
	    &
		\includegraphics[width=\resLen]{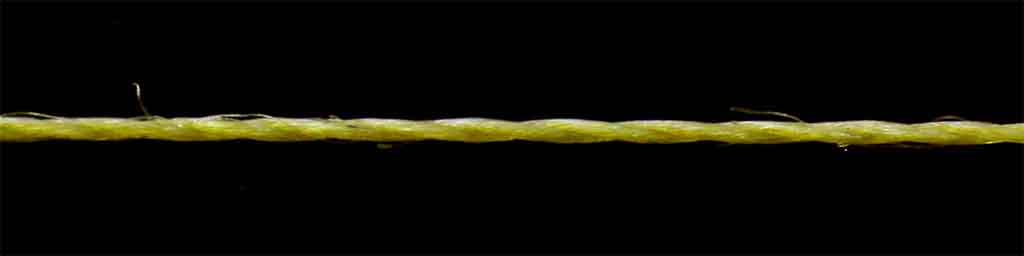}
		&
		\includegraphics[width=\resLen]{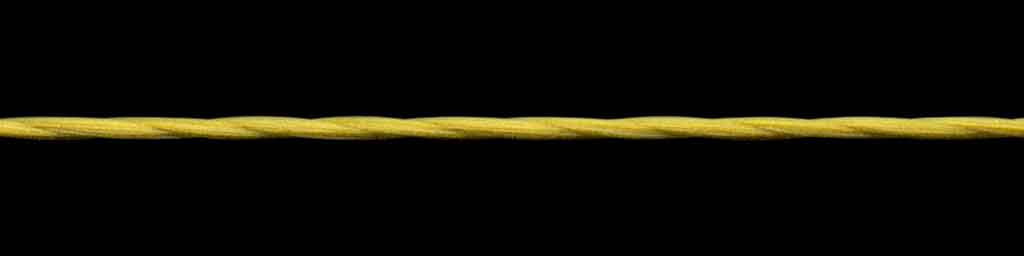}
		&
		\includegraphics[width=\resLen]{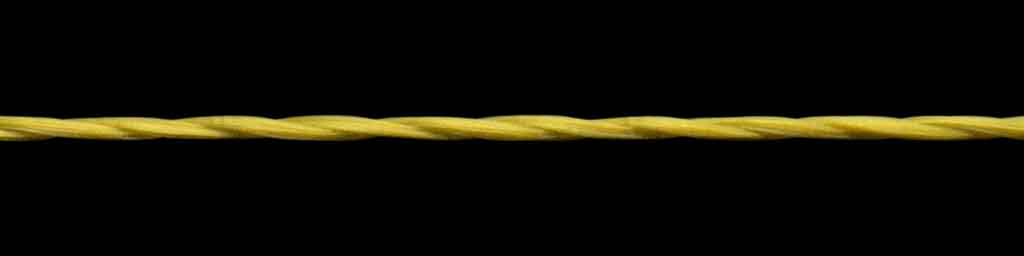}
		&
		\includegraphics[width=\resLen]{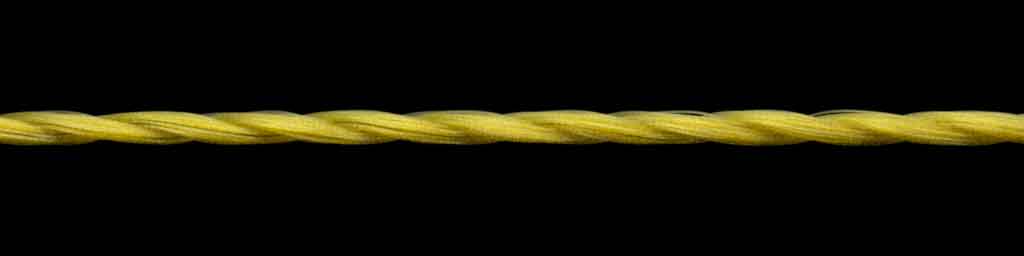}
		\\
		&
		\includegraphics[width=\resLen]{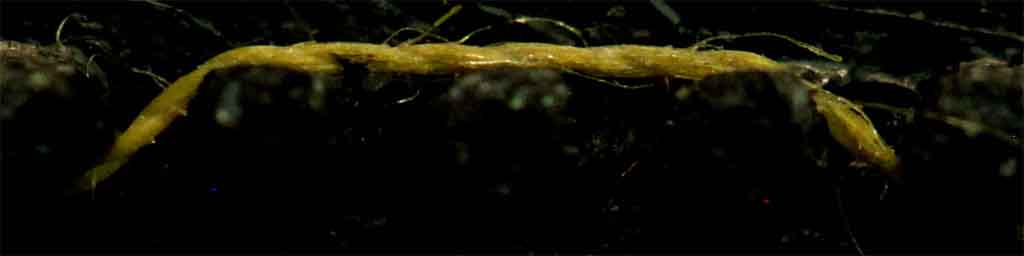}
		&
		\includegraphics[width=\resLen]{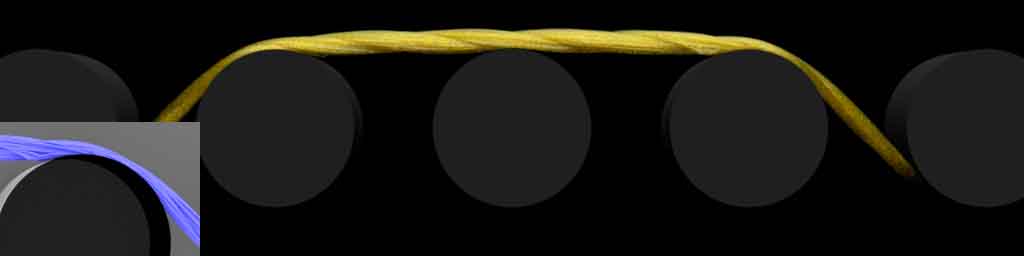}
		&
		\includegraphics[width=\resLen]{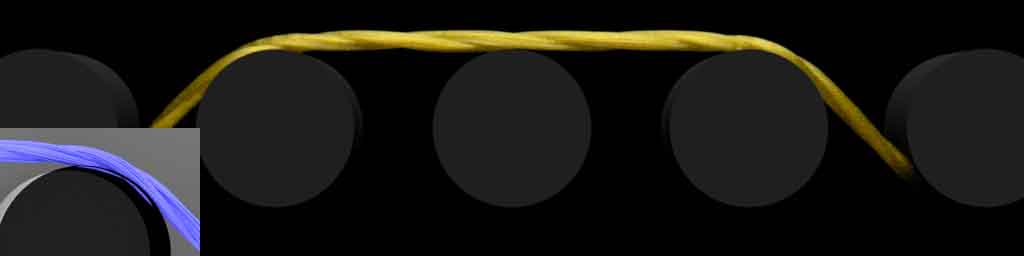}
		&
		\includegraphics[width=\resLen]{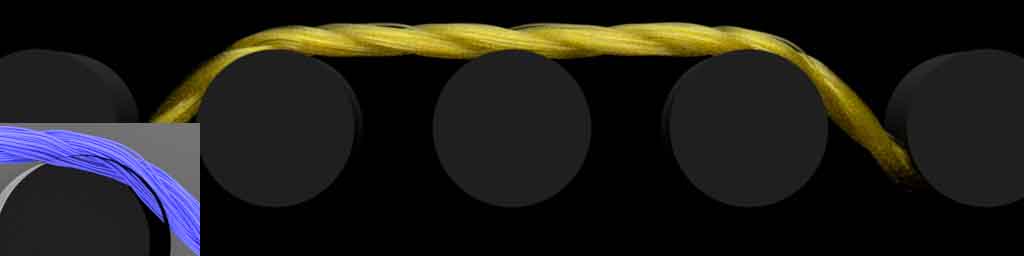}
		\\
		\multirow{2}{*}[13pt]{\rotatebox[origin=c]{90}{Cotton (3-ply)}}
		&
		\includegraphics[width=\resLen]{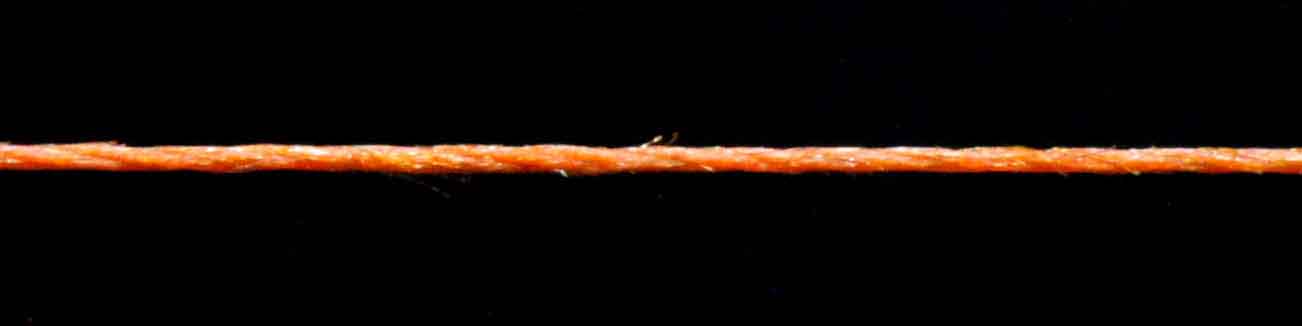}
		&
		\includegraphics[width=\resLen]{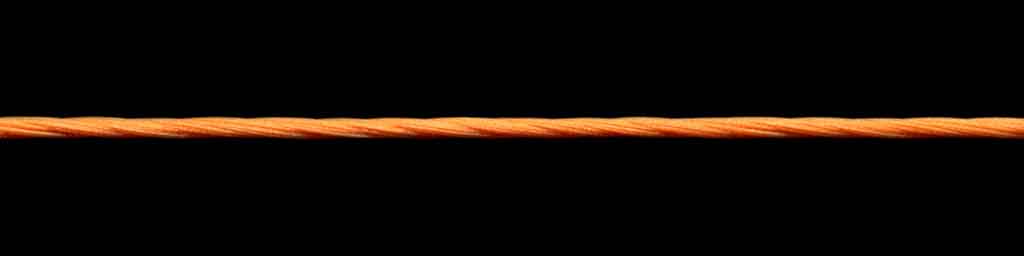}
		&
		\includegraphics[width=\resLen]{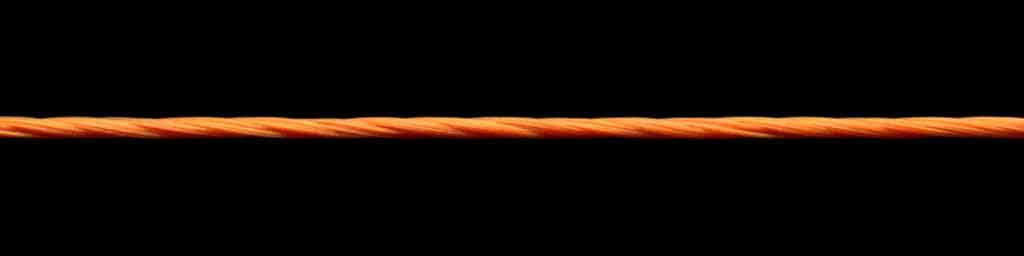}
		&
		\includegraphics[width=\resLen]{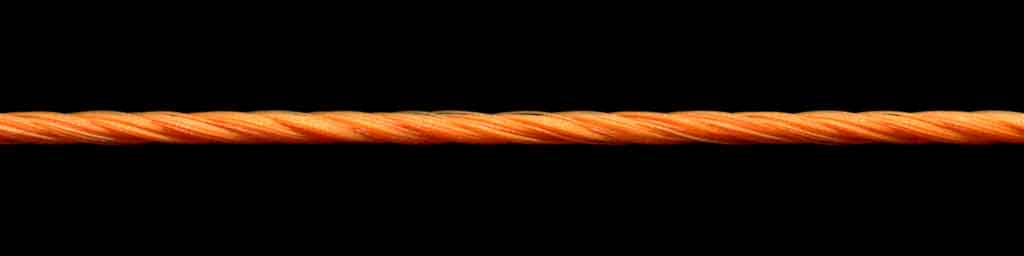}
		\\
		&
		\includegraphics[width=\resLen]{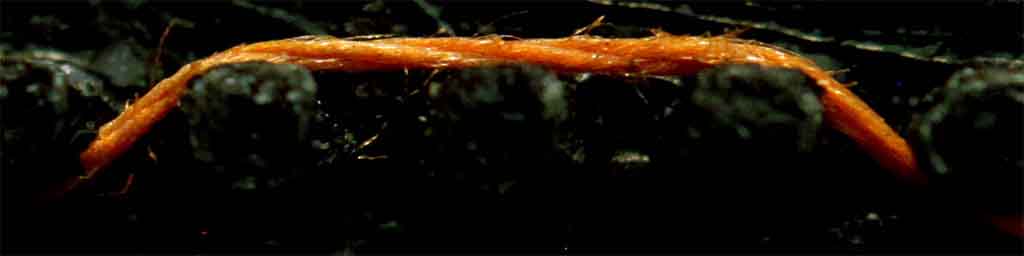}
		&
		\includegraphics[width=\resLen]{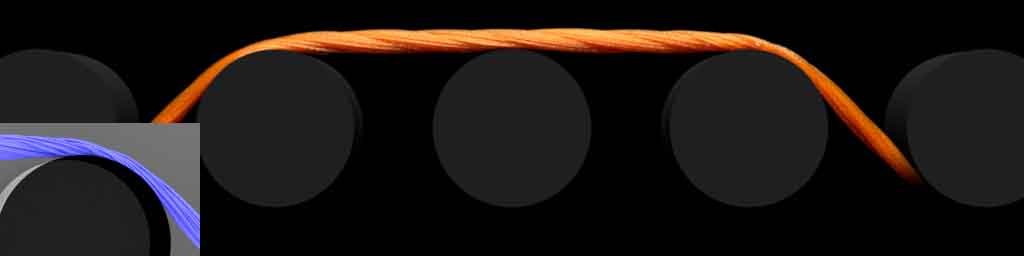}
		&
		\includegraphics[width=\resLen]{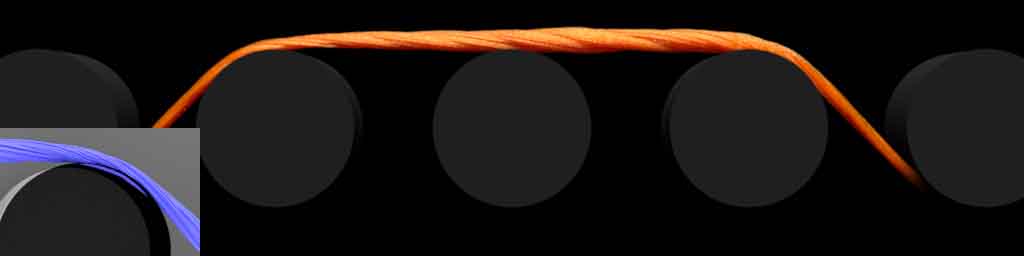}
		&
		\includegraphics[width=\resLen]{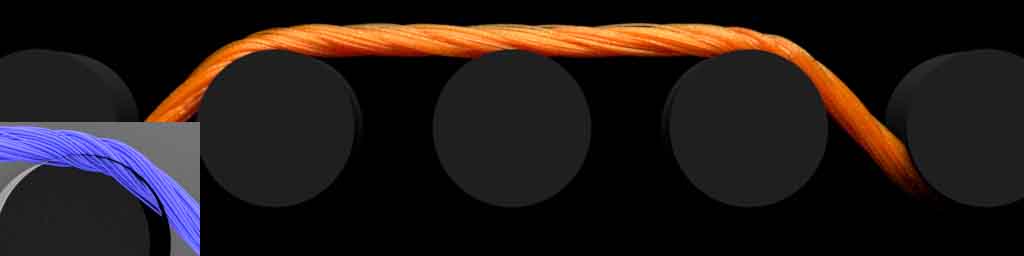}
		\\
		\multirow{2}{*}[13pt]{\rotatebox[origin=c]{90}{Rayon (2-ply)}}
		&
		\includegraphics[width=\resLen]{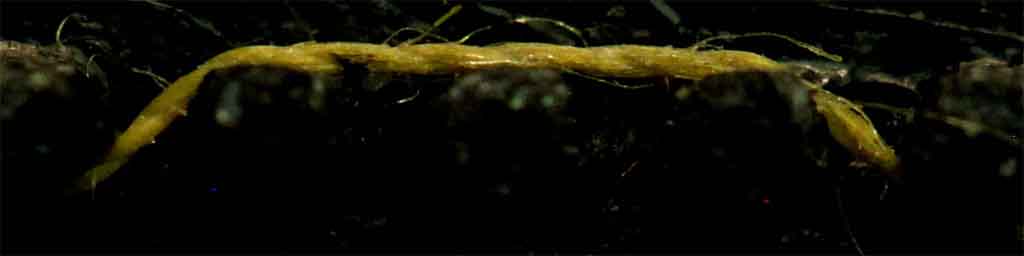}
		&
		\includegraphics[width=\resLen]{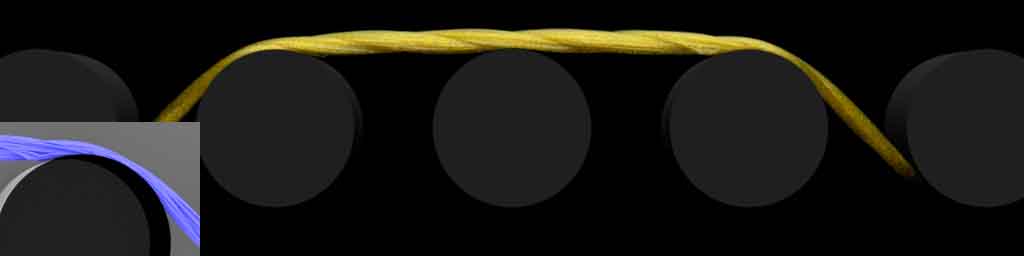}
		&
		\includegraphics[width=\resLen]{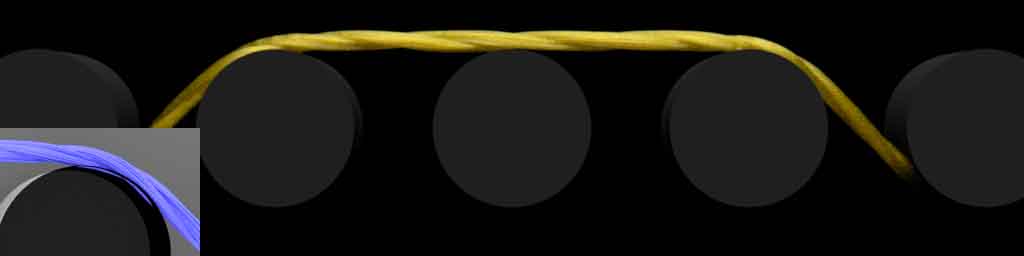}
		&
		\includegraphics[width=\resLen]{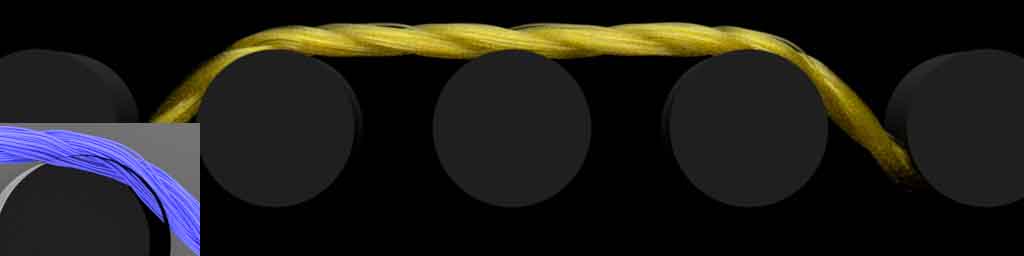}
		\\
		&
		\includegraphics[width=\resLen]{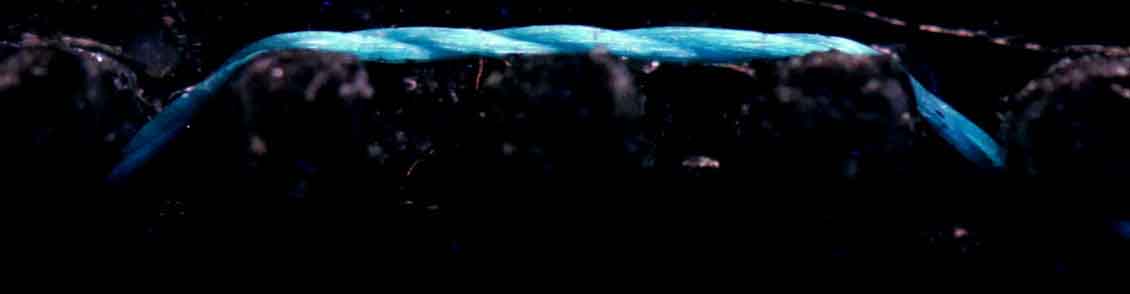}
		&
		\includegraphics[width=\resLen]{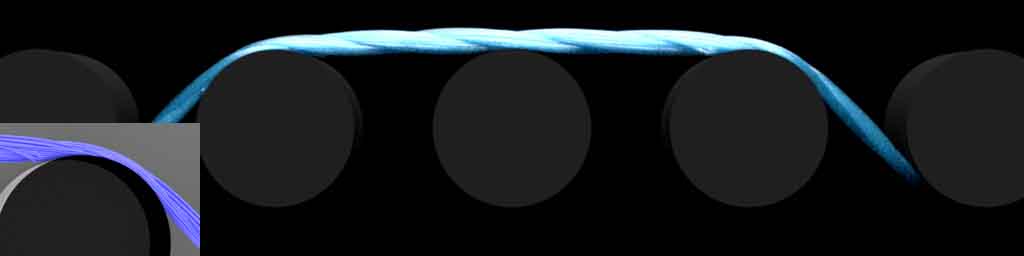}
		&
		\includegraphics[width=\resLen]{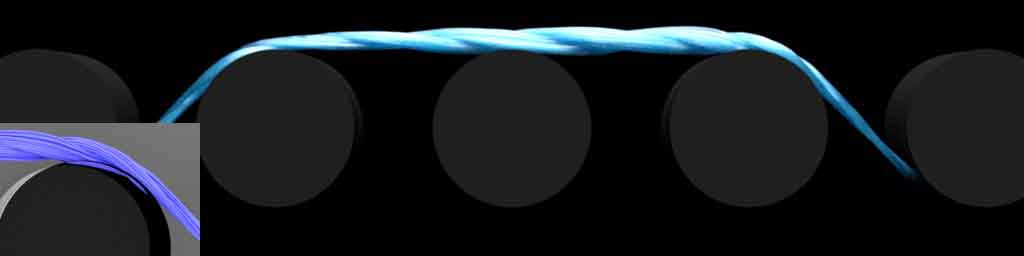}
		&
		\includegraphics[width=\resLen]{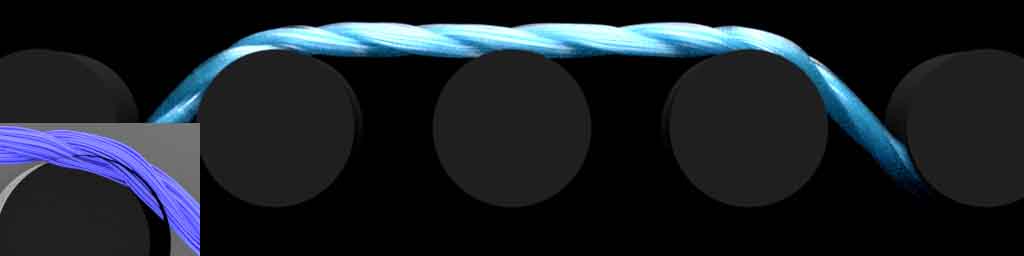}
		\\[3pt]
	\end{tabular}
	\setlength{\resLen}{0.162\textwidth}
	\begin{tabular}{cccccc}
	    (b2) Fiber-level sim.
	    &
	    (c2) Ours
	    &
	    (d2) Baseline
	    &
	    (b3) Fiber-level sim.
	    &
	    (c3) Ours
	    &
	    (d3) Baseline
	    \\
		\includegraphics[width=\resLen]{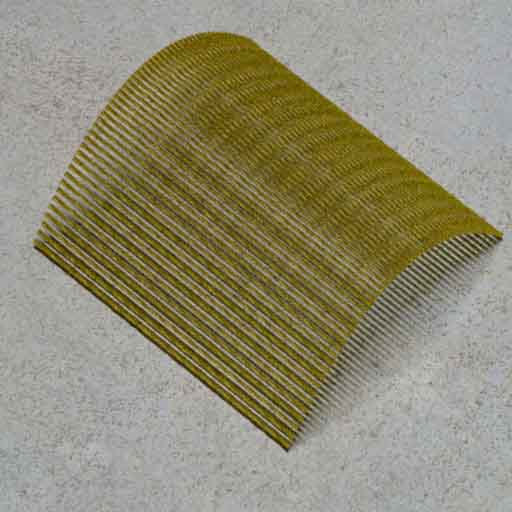}
		&
		\includegraphics[width=\resLen]{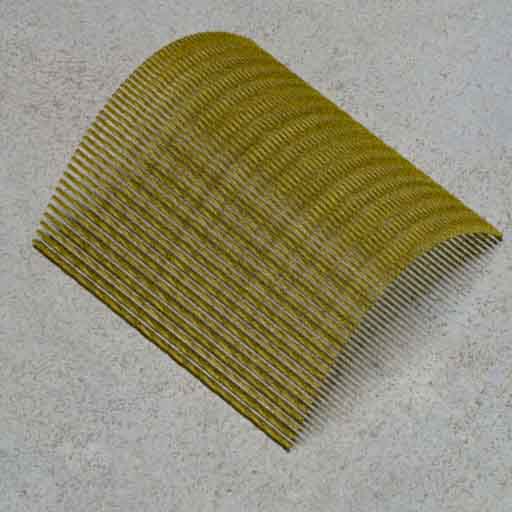}
		&
		\includegraphics[width=\resLen]{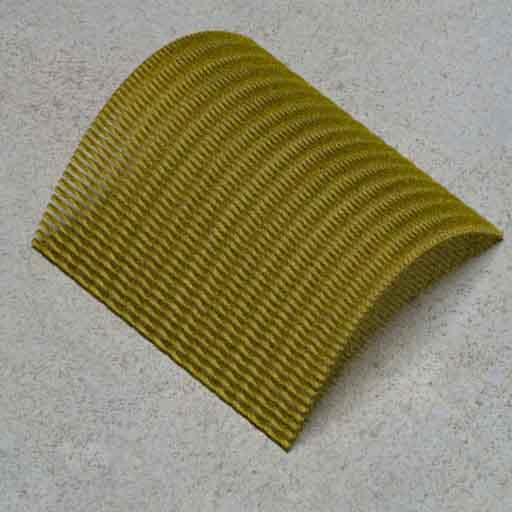}
		&
		\includegraphics[width=\resLen]{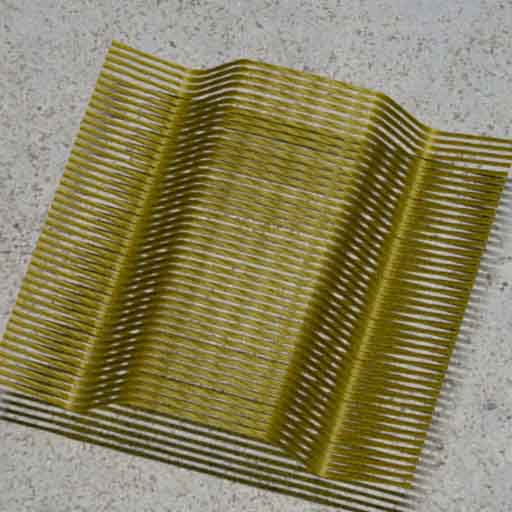}
		&
		\includegraphics[width=\resLen]{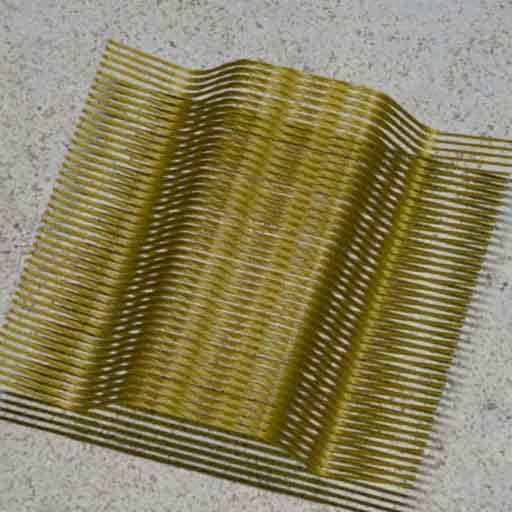}
		&
		\includegraphics[width=\resLen]{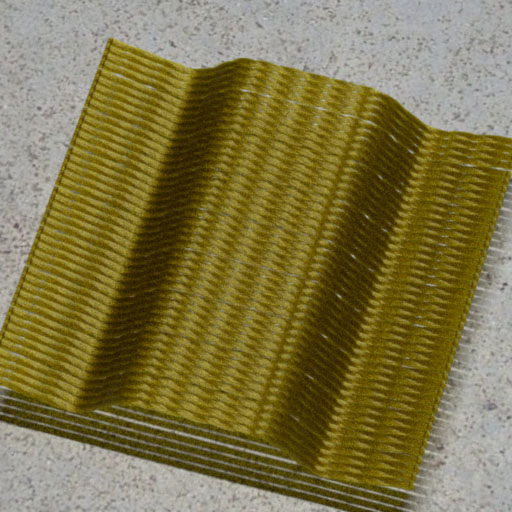}
		\\
		\includegraphics[width=\resLen]{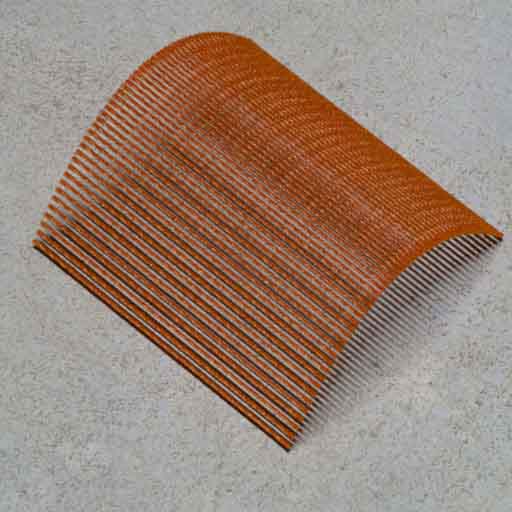}
		&
		\includegraphics[width=\resLen]{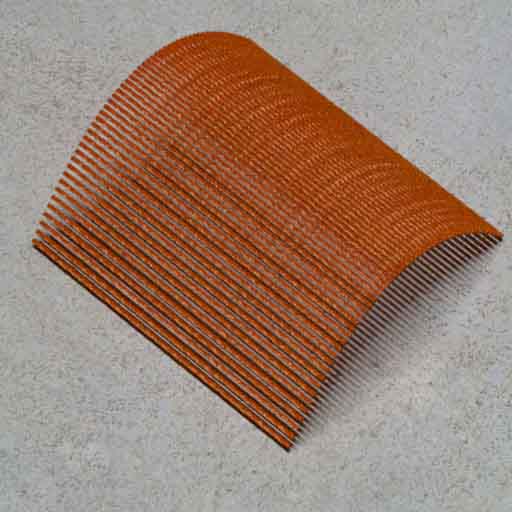}
		&
		\includegraphics[width=\resLen]{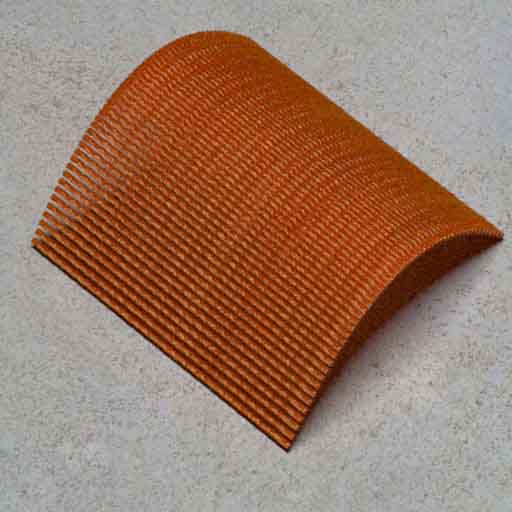}
		&
		\includegraphics[width=\resLen]{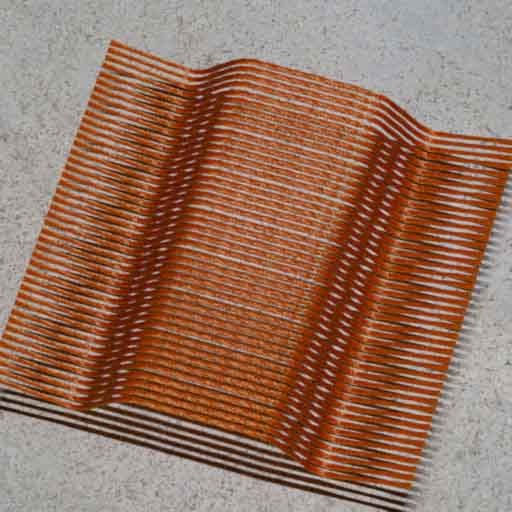}
		&
		\includegraphics[width=\resLen]{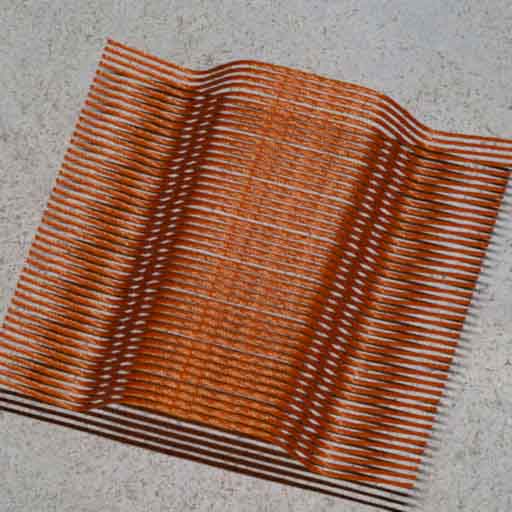}
		&
		\includegraphics[width=\resLen]{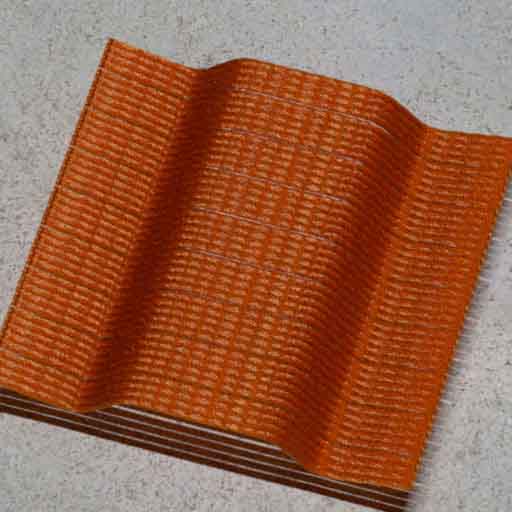}
		\\
	    \includegraphics[width=\resLen]{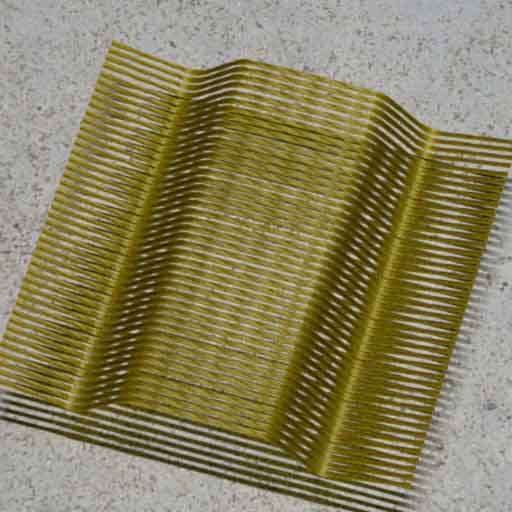}
		&
		\includegraphics[width=\resLen]{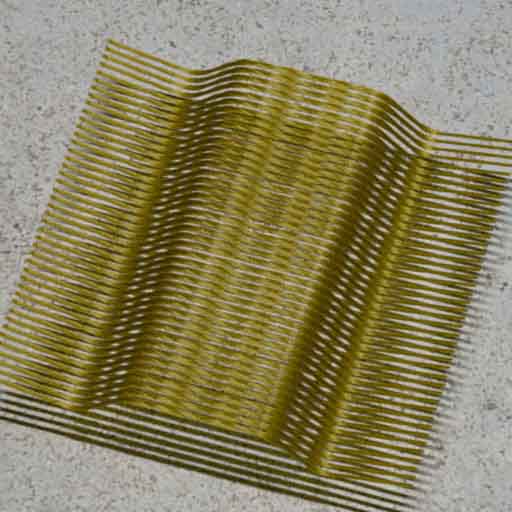}
		&
		\includegraphics[width=\resLen]{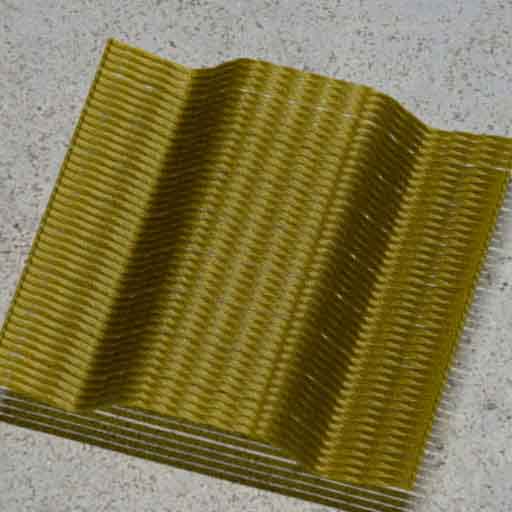}
		&
		\includegraphics[width=\resLen]{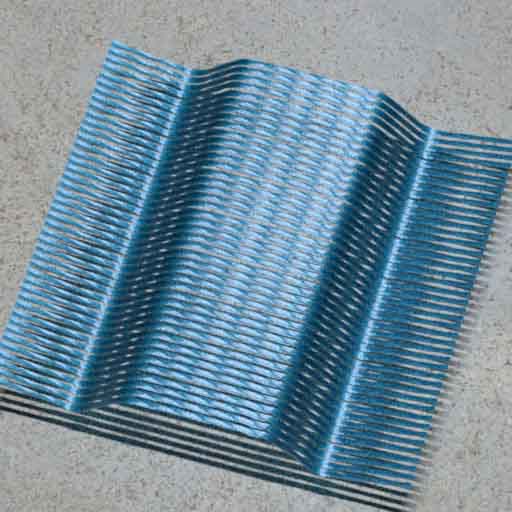}
		&
		\includegraphics[width=\resLen]{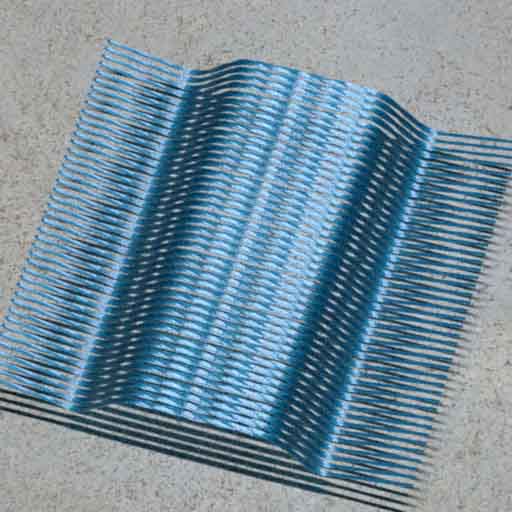}
		&
		\includegraphics[width=\resLen]{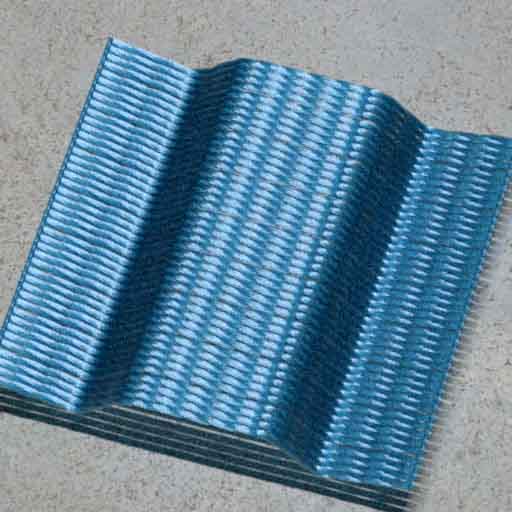}
	\end{tabular}
    \caption{\label{fig:result_single}
    	\textbf{Generated models for single yarns.}
        We demonstrate the effectiveness of our method by comparing our results~(c) to photographs~(a) and reference fiber-level simulations~(b) using three types of yarns: polyester, cotton, and rayon.
        Column~(d) shows procedurally generated yarns with fiber mechanics neglected.
        The top images~(1) show detailed fiber microstructures from individual yarns.
        The bottom images~(2 and 3) use tiled versions of the top results to demonstrate how difference in those structures affect macro-scale appearance.
    }
\end{figure*}

\paragraph*{\em Single-yarn results}
We validate the accuracy of our method by comparing to photographs and reference simulations at the fiber level.
In Figure~\ref{fig:result_single}, we show experimental results for three types of yarns, rayon, cotton, and polyester,
each using fitted procedural parameters from the previous work~\cite{Zhao:2016:FPY}.
We stretch all these yarns both directly and through a number of 3D-printed
cylinders. Both situations are considerably different from our training
simulations.
Our method is able to accurately estimate the deformation of the yarns'
fiber-level microstructures (Figure~\ref{fig:result_single}-c), which cannot be
closely captured using previous methods that neglects fiber mechanics
(Figure~\ref{fig:result_single}-d).  Accurate estimation of fiber
micro-geometry is crucial for producing fabric appearance at
larger scales, as demonstrated at the bottom of the figure.  Please refer to
the accompanying video for the corresponding animations.

This experiment also validates our parameter homogenization algorithm. The (b1) column of \figref{result_single}
is generated using fiber-level simulation, while the (c1) column is generated using yarn-level simulation with fitted parameters.
It is evident that the yarn centerlines resulted from yarn-level simulation with the fitted material parameters is able to closely match 
those from the fiber-level simulation.

\begin{figure}[t]
    \centering
    \small
    \setlength{\resLen}{0.45\columnwidth}
	\addtolength{\tabcolsep}{-15pt}
	\begin{tabular}{cc}
		\includegraphics[width=\resLen]{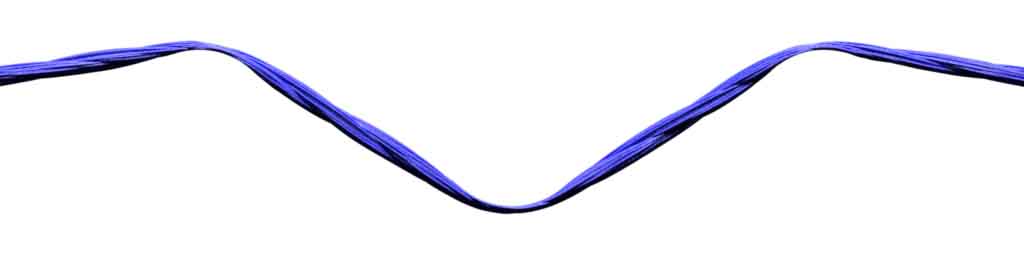}	&
		\includegraphics[width=\resLen]{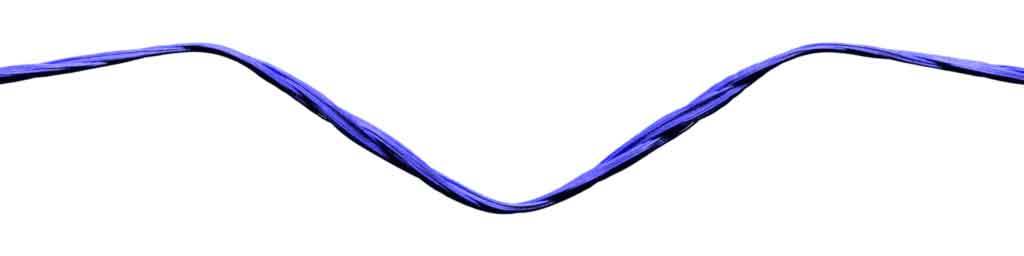}
		\\[-5pt]
		(a) Reference & (b) Regression neural network
		\\[2pt]
		\includegraphics[width=\resLen]{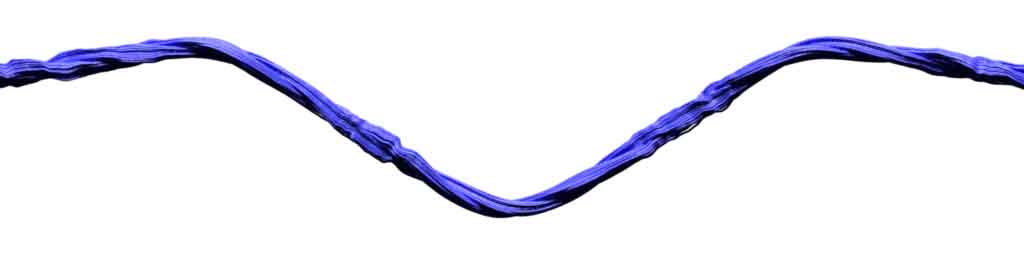} &
		\includegraphics[width=\resLen]{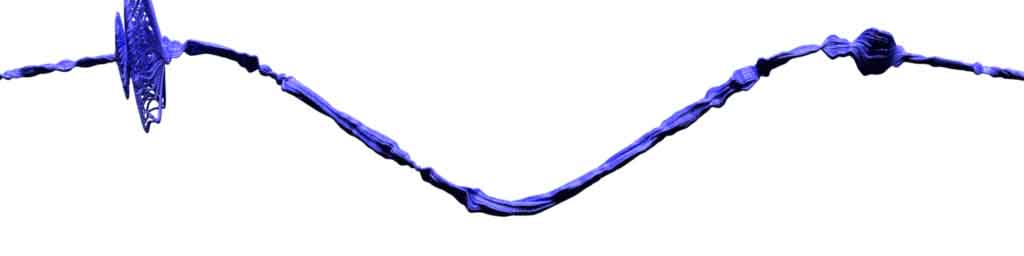}
		\\[-5pt]
		(c) Linear regression & (d) Polynomial regression (deg.-2)
		\\[2pt]
		\includegraphics[width=\resLen]{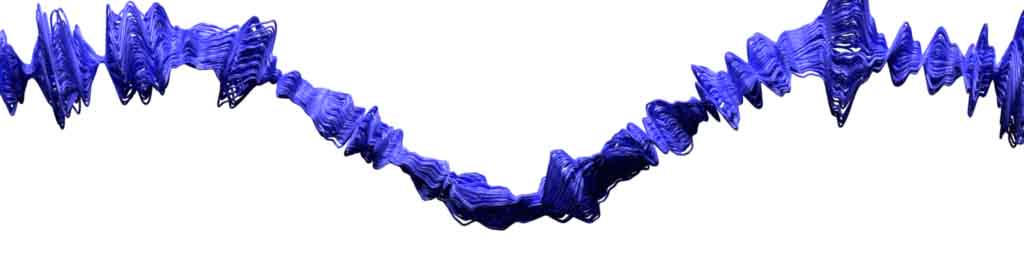}	&
		\includegraphics[width=\resLen]{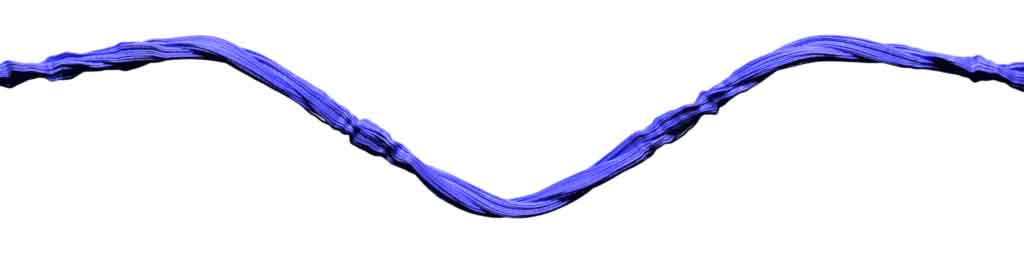}
		\\[-5pt]
		(e) Polynomial regression (deg.-4) & (f) Gaussian process
	\end{tabular}
    \caption{\label{fig:alter_reg}
        \textbf{Alternative regression methods} such as linear and polynomial regressions as well as Gaussian processes have difficulties in capturing the complex relations between simulated yarn-level deformation gradients and desired cross-sectional transformations of fiber centers.
    }
\end{figure}

\paragraph*{\em Alternative regression models}
To learn the mapping between yarn-level simulated deformation gradients and desired transformations of fiber center, we chose to use a regression neural network due to the complexity of this mapping.
\figref{alter_reg} demonstrates the performance of several simpler regression models include linear and polynomial regression and Gaussian process.
Trained with the same data, our regression network behaves much more stably and provides results with superior visual quality.

\paragraph*{\em Multi-yarn results} 
\figref{result_6x6} shows a comparison between the reference (obtained via fiber-level simulations) and our result. Because explicitly simulating individual fibers is intractable at a large scale, we performed the simulation for a small patch with $4 \times 4$ yarns and tiled the resulting fiber geometries for both the reference (i.e., directly simulating at fiber level) and our method (i.e., enriching the yarn-level simulation) to provide a macroscopic comparison.
As shown in the figure, our result resembles the reference at both micro and macro scales, while previous methods fail to capture the yarns' thinning effect (\figref{result_6x6}-c).

Lastly, we perform a qualitative comparison between our results and photographs (\figref{photo_comp}).
In this experiment, we take photographs of a real knitted fabric under rest and stretched states.
Additionally, we create a virtual model after the physical sample and tweaked the optical parameters to match the appearance in rest states (\figref{photo_comp}-a1 and b1).
Then, we simulate a stretching of this model at the yarn level and apply our technique to the stretched model.
Our method (\figref{photo_comp}-b2) successfully predicts the color change caused by yarn and fiber deformations.
Please refer to the accompanying video for an animated version of this comparison.

\begin{table}[t]
	\scriptsize
	\caption{\label{tab:performance}
		\textbf{Simulation and processing time statistics.}
		In Figure~\protect\ref{fig:result_single}, fiber-level simulations are used only to generate references.
		Numbers in the right-most column indicate the total time spent on procedurally generating fiber curves and deforming them by evaluating our regression network. The timing statistics are collected on a workstation with four Intel Xeon E5-2620 v3 CPUs with six cores each running at 2.40GHz.
	}
	\vspace{2mm}
	\renewcommand{\arraystretch}{1.3}
	\addtolength{\tabcolsep}{-6pt}
	\resizebox{\columnwidth}{!}{%
	\begin{tabular}{cccccc}
		\whline{0.8pt}
		& \textbf{Scene} & \textbf{ \# vertices } & \textbf{\thead{Fiber-level\\ simulation} } & \textbf{\thead{Yarn-level\\ simulation} }  & \textbf{\thead{Fiber generation\\+ NN evaluation} } \\
		\whline{0.05pt}
		& Figure~\protect\ref{fig:result_single} & 300 & 8 h & 15 min & 20 sec\\
		\whline{0.05pt}
		& Figure~\protect\ref{fig:result_knitted}-a & 85832 & - & 20 h & 1  min\\
		& Figure~\protect\ref{fig:result_knitted}-b & 85832 & - & 18 h & 30 sec\\
		& Figure~\protect\ref{fig:result_knitted}-c & 58545 & - & 28 h & 1  min\\
		\whline{0.05pt}
		& Figure~\protect\ref{fig:result_woven}-a & 100000 & - & 9  h & 1  min\\
		& Figure~\protect\ref{fig:result_woven}-b & 100000 & - & 10 h & 1  min\\
		& Figure~\protect\ref{fig:result_woven}-c & 1266000 & - & 110 h & 10 min\\
		\whline{0.8pt}
	\end{tabular}
}
\end{table}

\begin{figure}[t]
\centering
	\addtolength{\tabcolsep}{-4pt}
	\begin{tabular}{ccc}
		\begin{overpic}[width=0.15\textwidth]{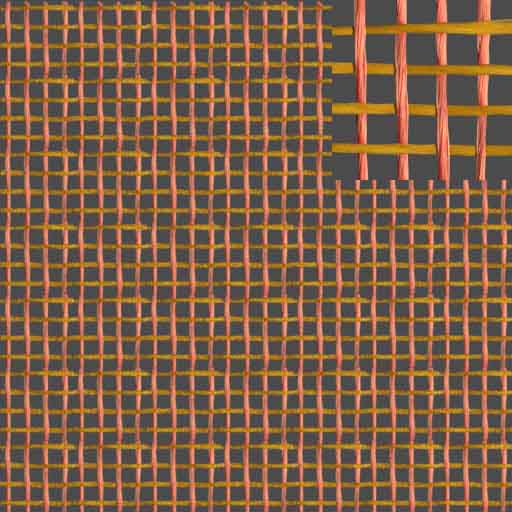}
			\put(2, 3){\small \color{white} \bfseries Reference}
		\end{overpic}
		&
		\begin{overpic}[width=0.15\textwidth]{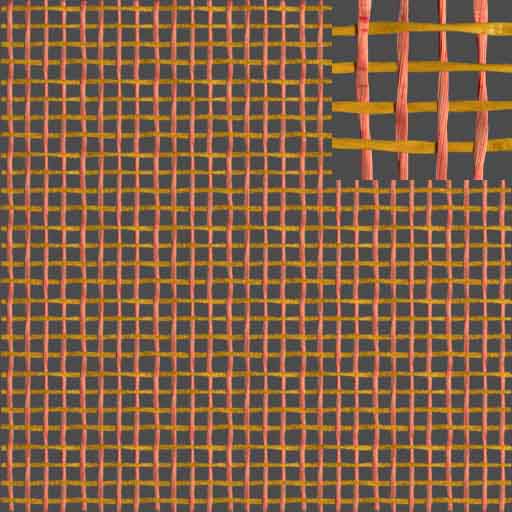}
			\put(2, 3){\small \color{white} \bfseries Ours}
		\end{overpic}
		&
		\begin{overpic}[width=0.15\textwidth]{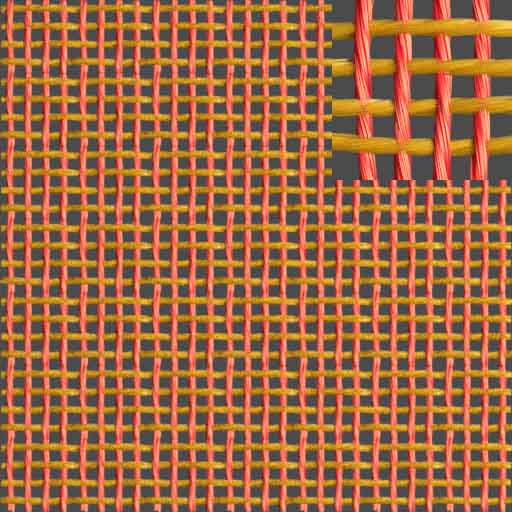}
			\put(2, 3){\small \color{white} \bfseries Baseline}
		\end{overpic}
	\end{tabular}
	\caption{\label{fig:result_6x6}
		\textbf{Generated models for a small textile.}
	    We simulate the dynamics of eight fibers (four horizontal and four
	    vertical), as shown in the insets. Our result generated by yarn-level
	    simulation and the fiber-deforming neural network matches closely the reference
	    fully simulated at the fiber level.
	}
\end{figure}

\subsection{Simulated Results}
\label{ssec:res_main}

We now demonstrate the effectiveness of our technique via a number of experiments.
Please see Table~\ref{tab:performance} for detailed statistics and performance numbers.

\begin{figure*}[p]
	\centering
	\setlength{\resLen}{1.98in}
	\addtolength{\tabcolsep}{-4pt}
	\vspace{0.1in}
	\begin{tabular}{cccc}
		\raisebox{3\height}{\rotatebox[origin=c]{90}{Rest shape}}
		&
		\begin{overpic}[height=\resLen]{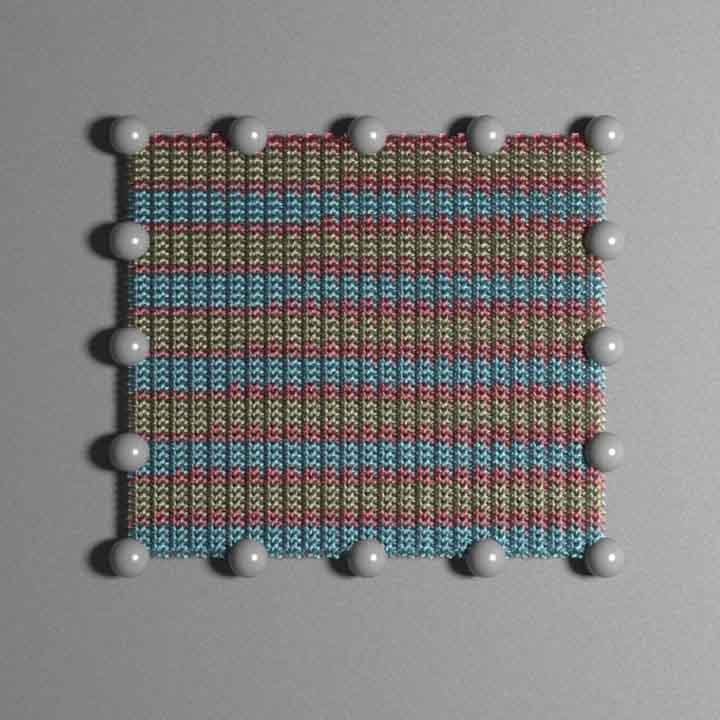}
			\put(2, 3){\small \color{white} \bfseries (a1)}
		\end{overpic}
		&
		\begin{overpic}[height=\resLen]{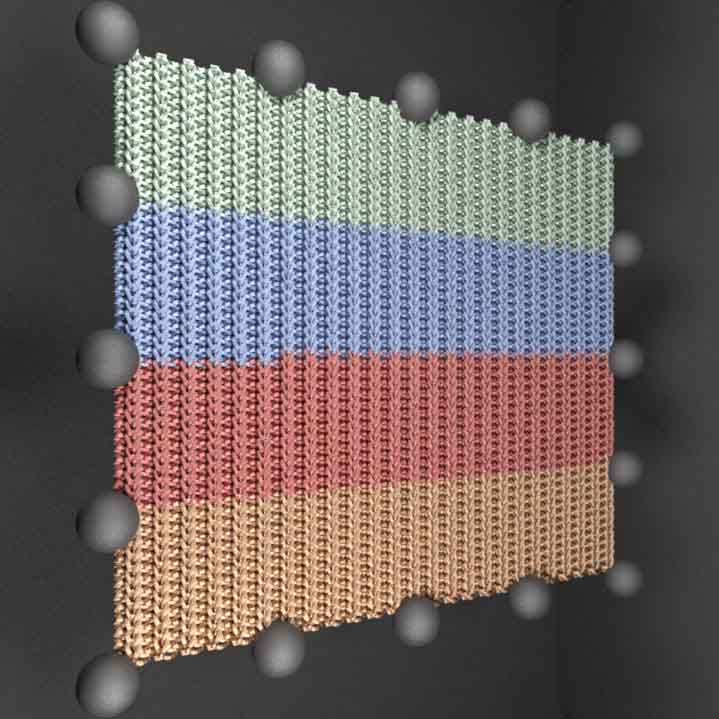}
			\put(2, 3){\small \color{white} \bfseries (b1)}
		\end{overpic}
		&
		\begin{overpic}[height=\resLen]{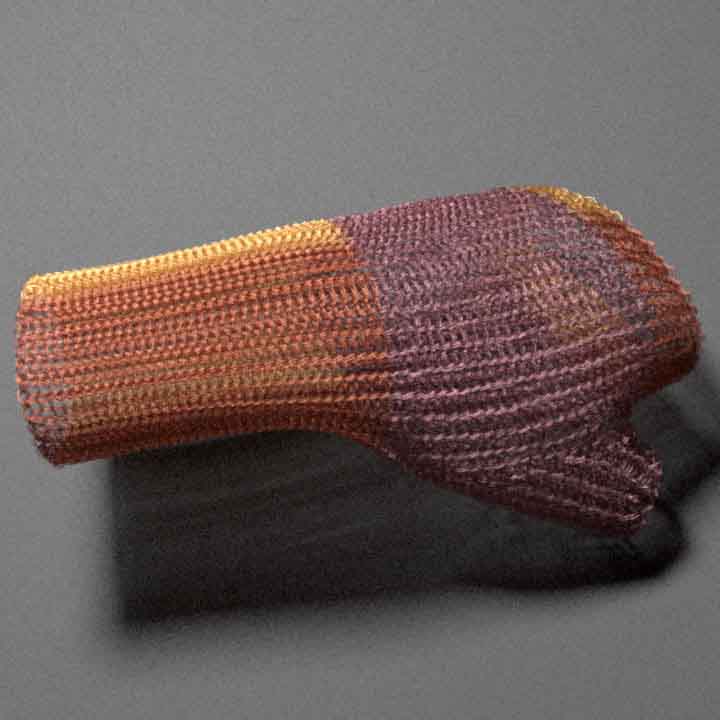}
			\put(2, 3){\small \color{white} \bfseries (c1)}
		\end{overpic}
	\end{tabular}
	\\[0.1in]
	\begin{tabular}{cc}
		Ours
		&
		Baseline
		\\
		\begin{overpic}[height=\resLen]{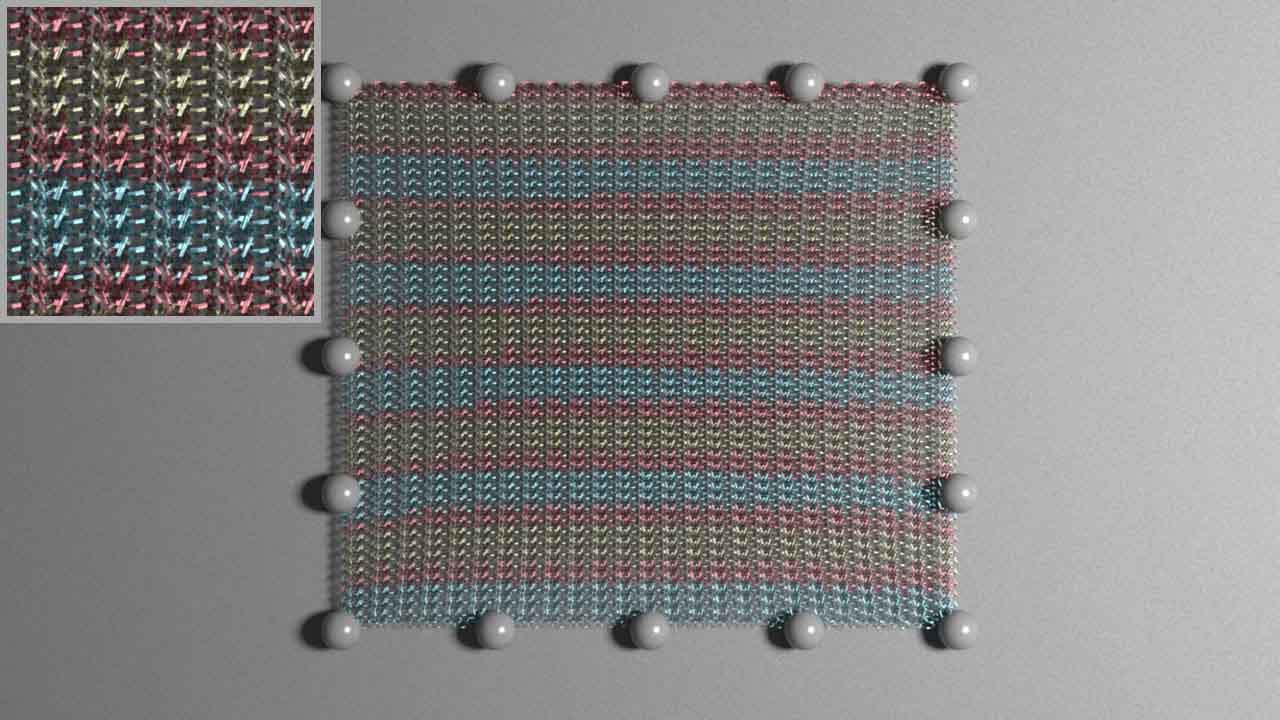}
			\put(2, 3){\small \color{white} \bfseries (a2)}
		\end{overpic}
		&
		\begin{overpic}[height=\resLen]{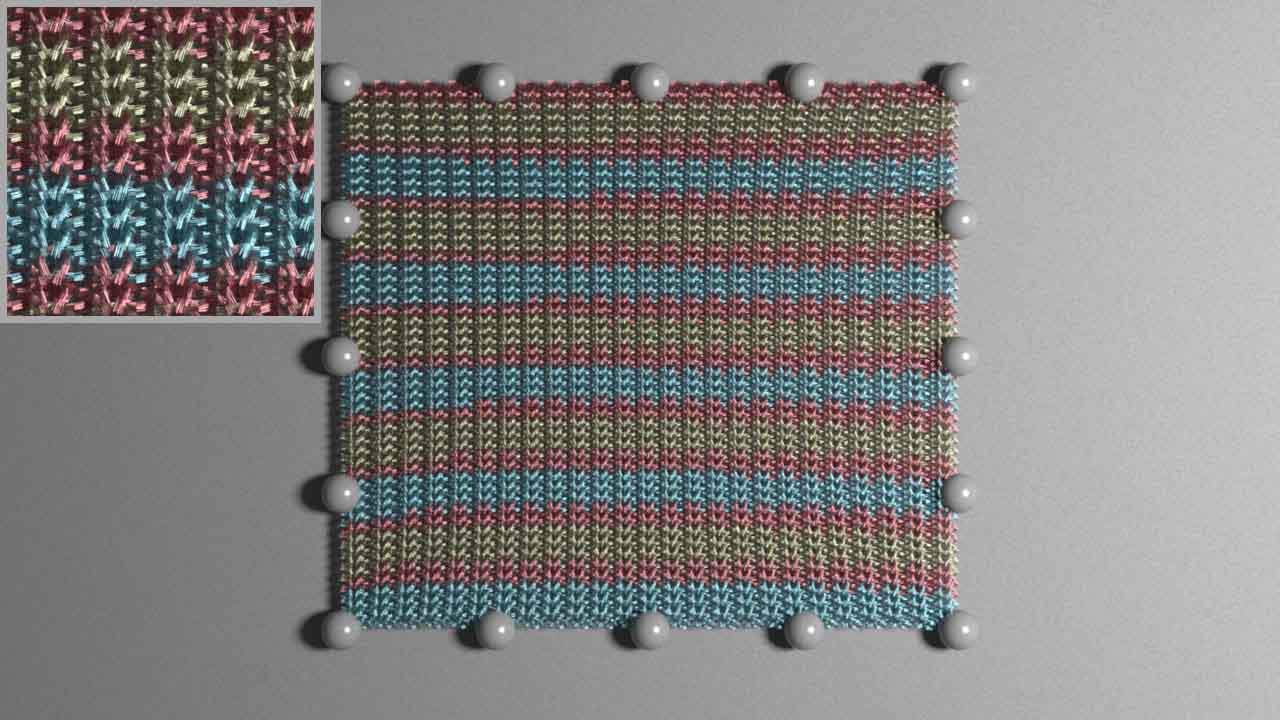}
			\put(2, 3){\small \color{white} \bfseries (a3)}
		\end{overpic}
		\\[5pt]
		
		\begin{overpic}[height=\resLen]{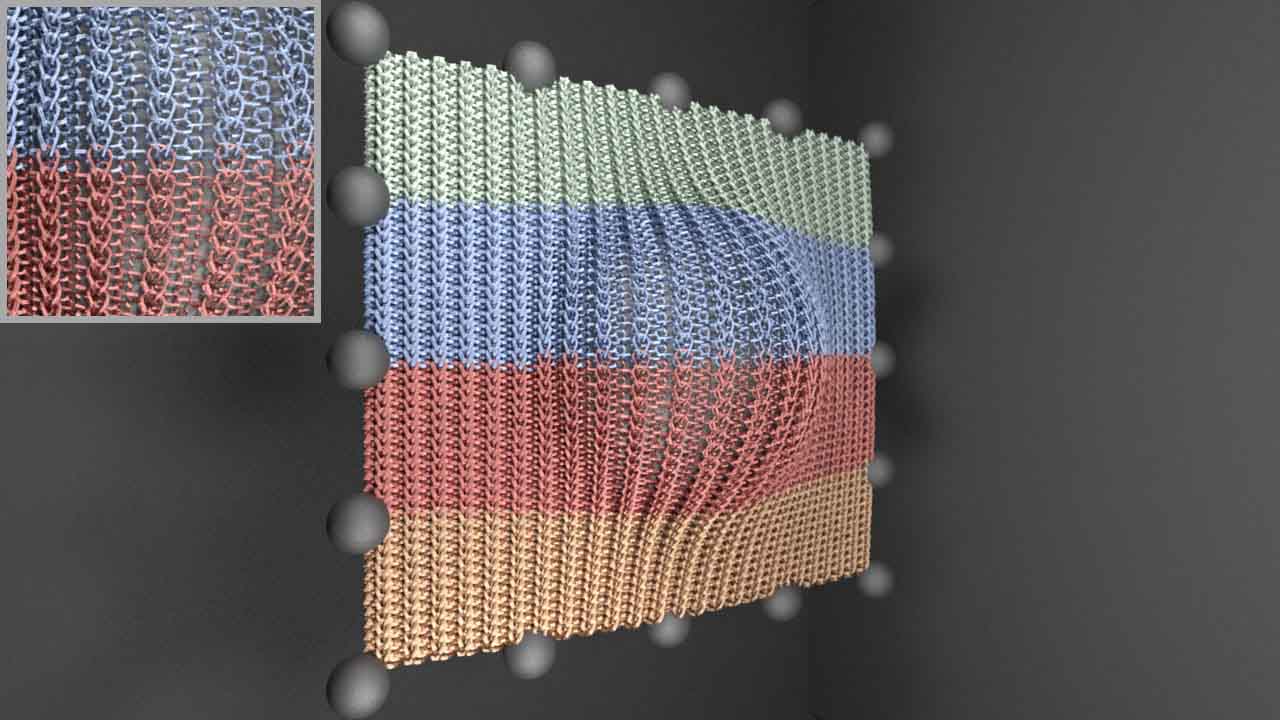}
			\put(2, 3){\small \color{white} \bfseries (b2)}
		\end{overpic}
		&
		\begin{overpic}[height=\resLen]{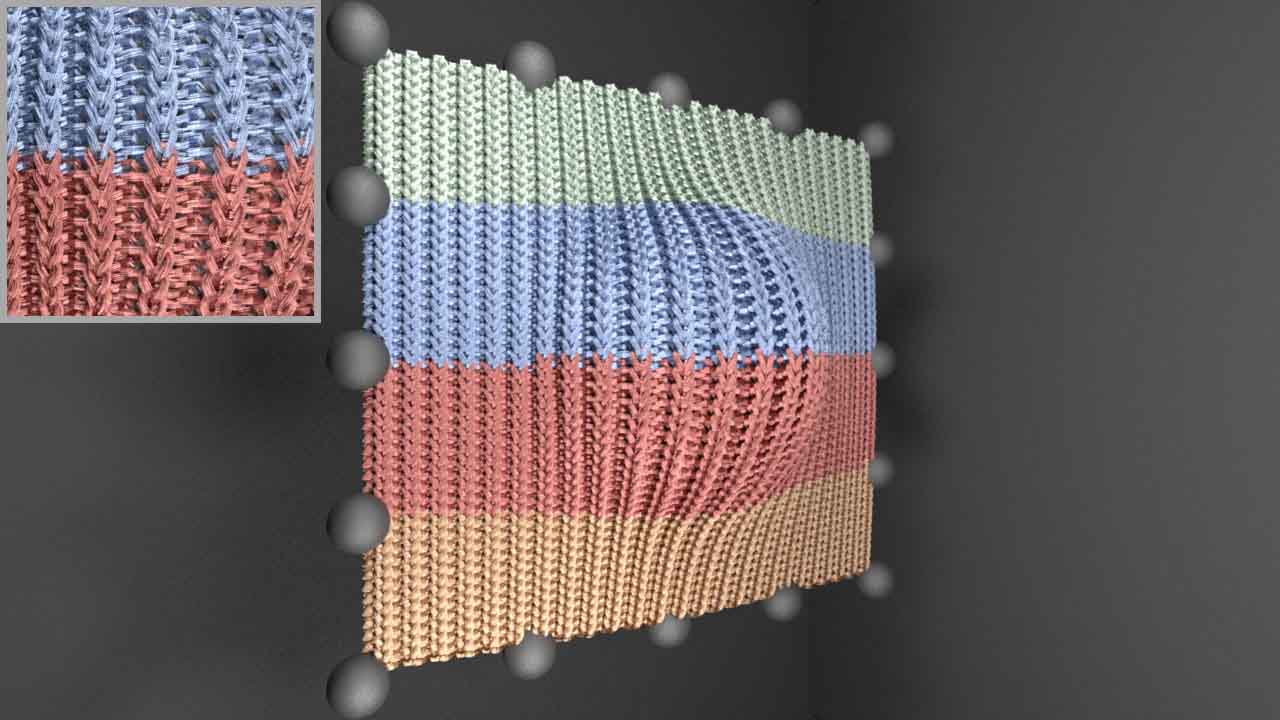}
			\put(2, 3){\small \color{white} \bfseries (b3)}
		\end{overpic}
		\\[5pt]
		\begin{overpic}[height=\resLen]{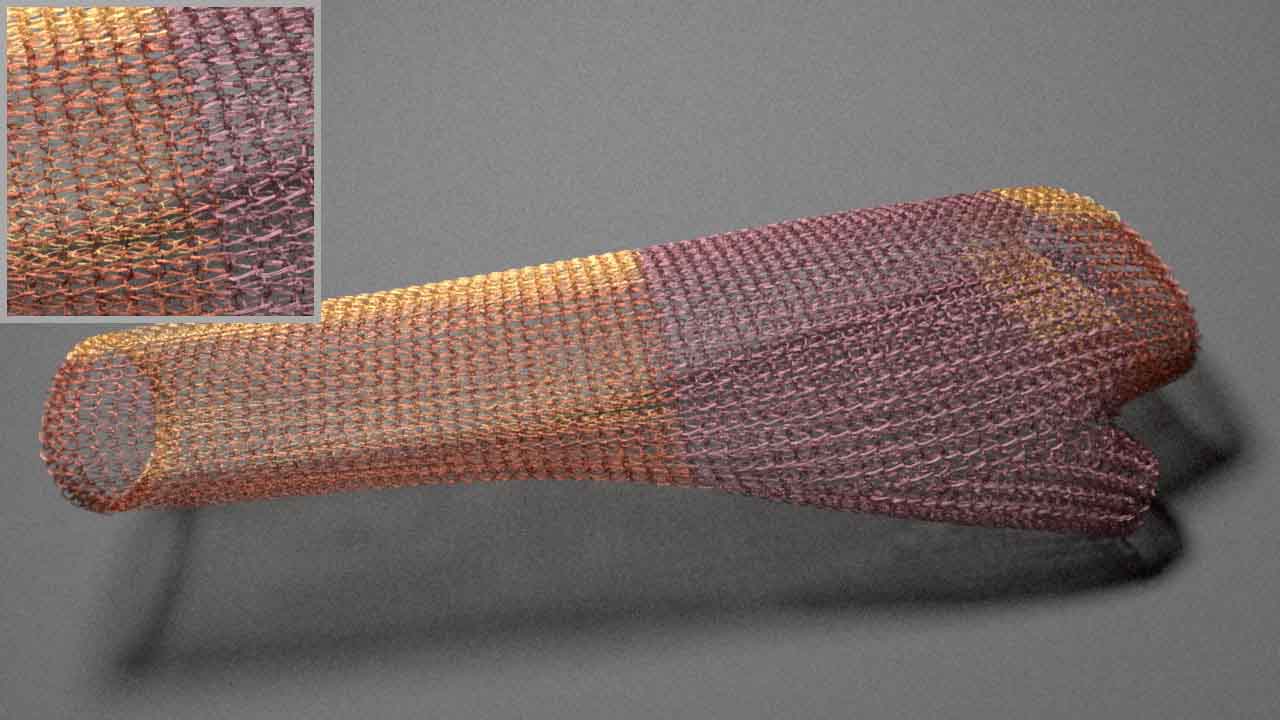}
			\put(2, 3){\small \color{white} \bfseries (c2)}
		\end{overpic}
		&
		\begin{overpic}[height=\resLen]{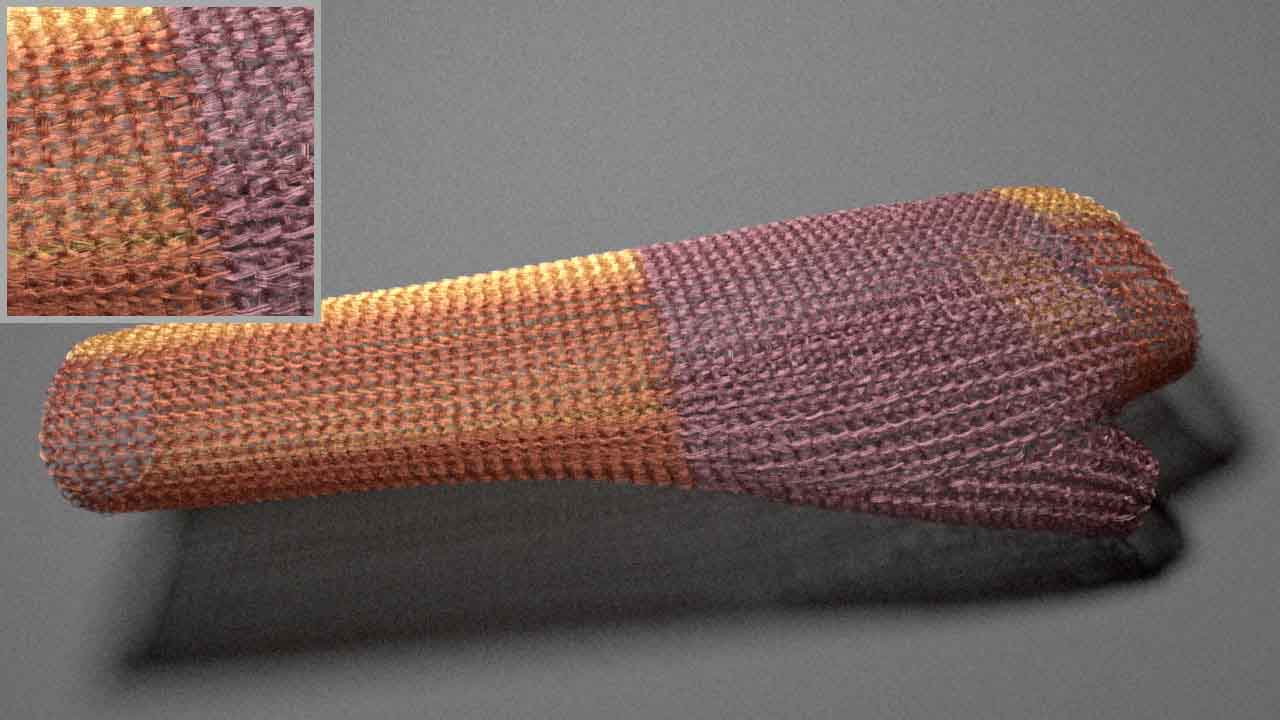}
			\put(2, 3){\small \color{white} \bfseries (c3)}
		\end{overpic}
	\end{tabular}
    \caption{\label{fig:result_knitted} 
    	\textbf{Generated models for knitted fabrics.}
    	Our technique can also capture the change of appearance of knitted fabrics caused by their mechanical responses to external forces.
    	(1)~fabrics in their rest shapes;
    	(2)~deformed fabrics modeled with our techniques;
    	(3)~deformed fabrics with identical yarn centerlines but with fiber-level deformations neglected.
    	The knitted fabrics in (a), (b), and (c) are comprised of cotton, polyester and rayon yarns, respectively.
    }
\end{figure*}

Our technique can be applied independently to individual yarns of a full fabric.
\figref{result_knitted} shows rendered results of full knitted textiles.
\figref{result_knitted}-a shows a knitted fabric with a slip-stitch rib pattern being stretched outward, causing the entire fabric to become more see-through.
In \figref{result_knitted}-b, we show a rigid sphere being pushed toward another knitted fabric with the same pattern.
The contact force applied by the sphere varies across the surface of this fabric, yielding spatially varying yarn and fiber deformations.
Our model is able to reproduce this effect in a visually convincing manner.
In \figref{result_knitted}-c, we show the stretching of a knitted glove modeled after the work by Wu and Yuksel~\cite{Wu2017}.
The mechanical responses of the yarns and fibers in this glove change its macro-scale appearance drastically: the glove not only becomes more see-through but also appears shinier due to the fibers being better aligned.
Our technique successfully captures this complex change of appearance without require simulating individual cloth fibers.

\begin{figure*}[p]
	\centering
    \setlength{\resLen}{1.98in}
	\addtolength{\tabcolsep}{-4pt}
	\vspace{0.1in}
	\begin{tabular}{cccc}
		\raisebox{3\height}{\rotatebox[origin=c]{90}{Rest shape}}
		&
		\begin{overpic}[height=\resLen]{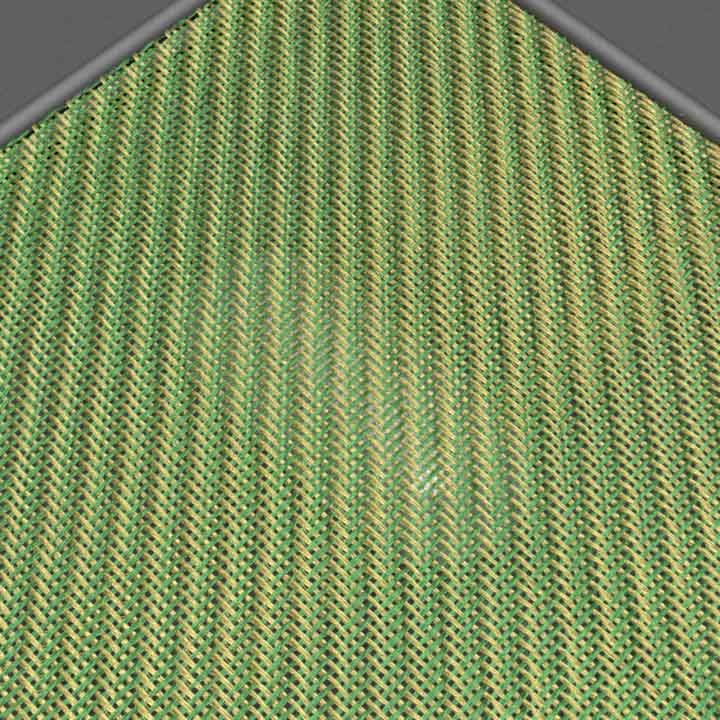}
			\put(2, 3){\small \color{white} \bfseries (a1)}
		\end{overpic}
		&
		\begin{overpic}[height=\resLen]{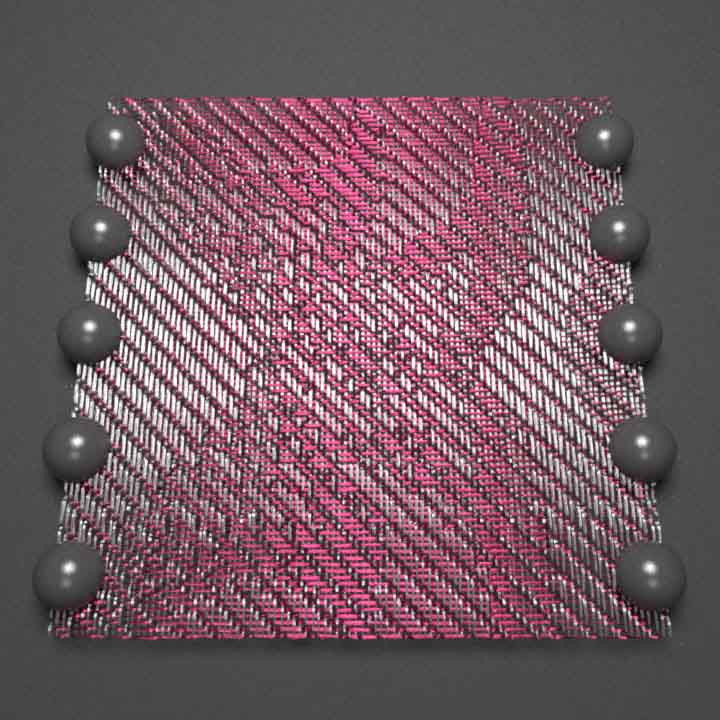}
			\put(2, 3){\small \color{white} \bfseries (b1)}
		\end{overpic}
		&
		\begin{overpic}[height=\resLen]{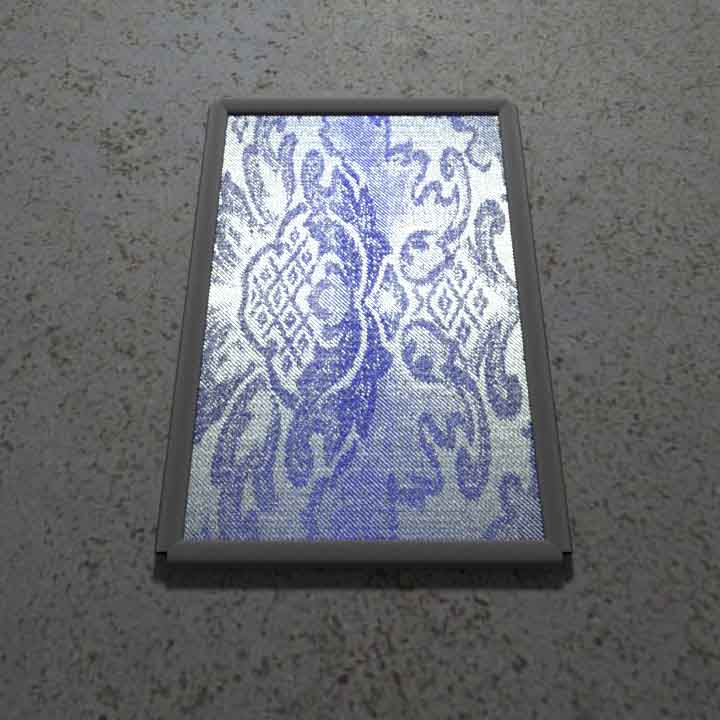}
			\put(2, 3){\small \color{white} \bfseries (c1)}
		\end{overpic}
	\end{tabular}
	\\[0.1in]
	\begin{tabular}{cc}
		Ours 
		&
		Baseline
		\\
		\begin{overpic}[height=\resLen]{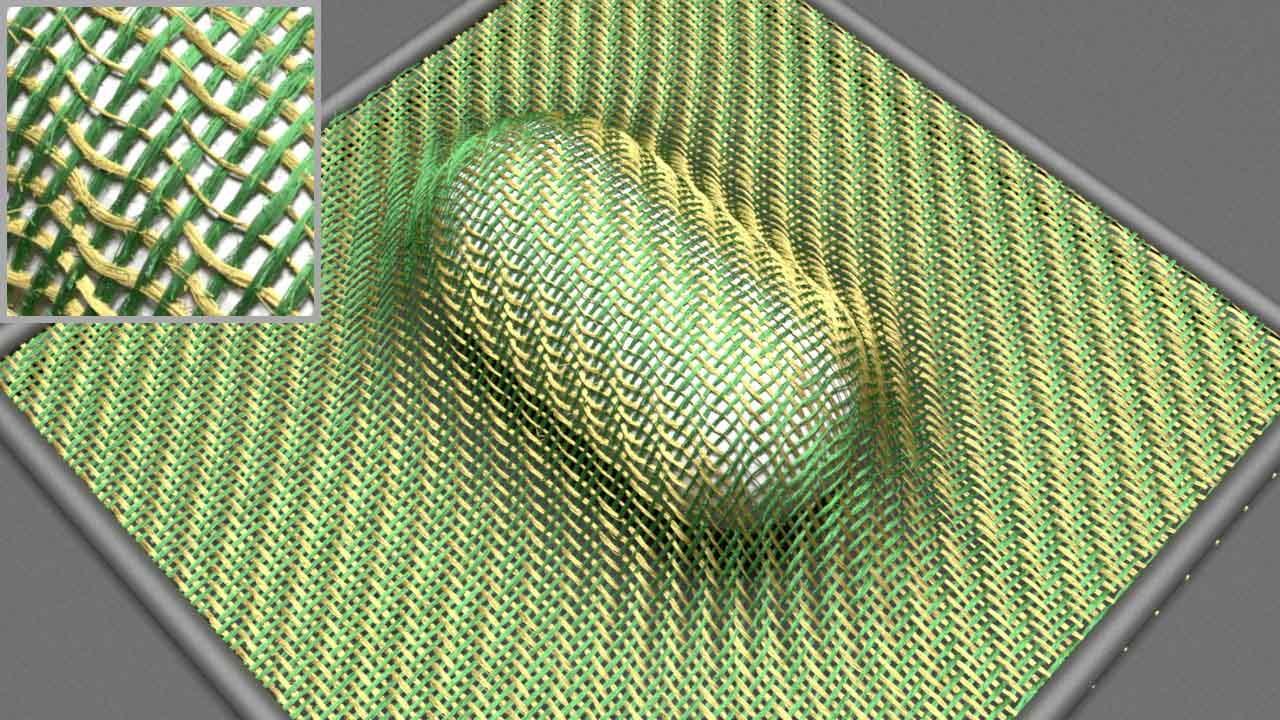}
			\put(2, 3){\small \color{white} \bfseries (a2)}
		\end{overpic}
		&
		\begin{overpic}[height=\resLen]{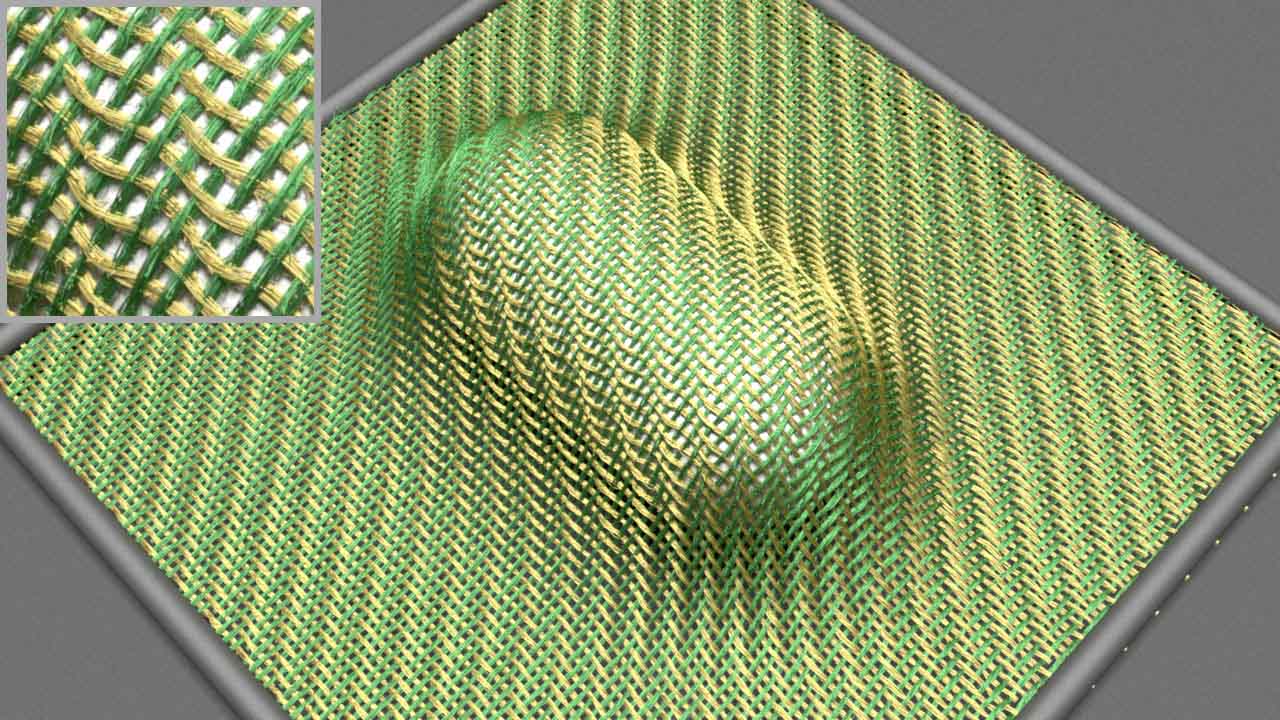}
			\put(2, 3){\small \color{white} \bfseries (a3)}
		\end{overpic}
		\\[5pt]
		\begin{overpic}[height=\resLen]{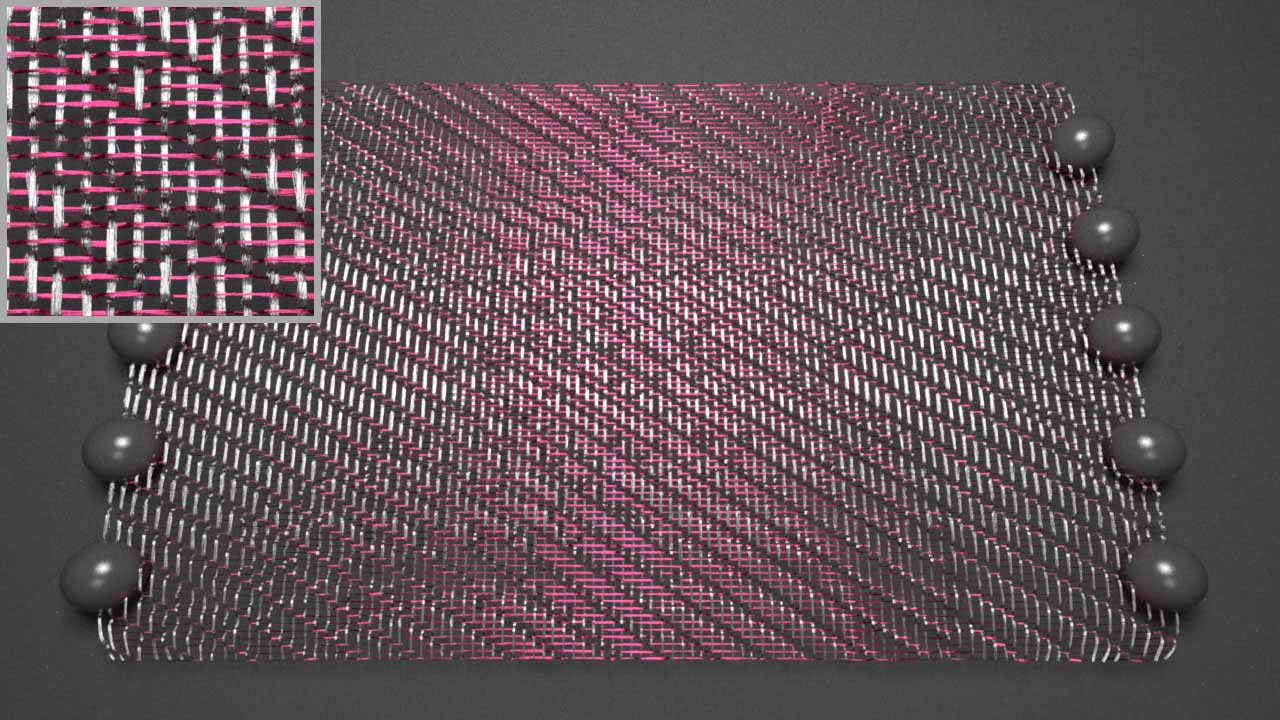}
			\put(2, 3){\small \color{white} \bfseries (b2)}
		\end{overpic}
		&
		\begin{overpic}[height=\resLen]{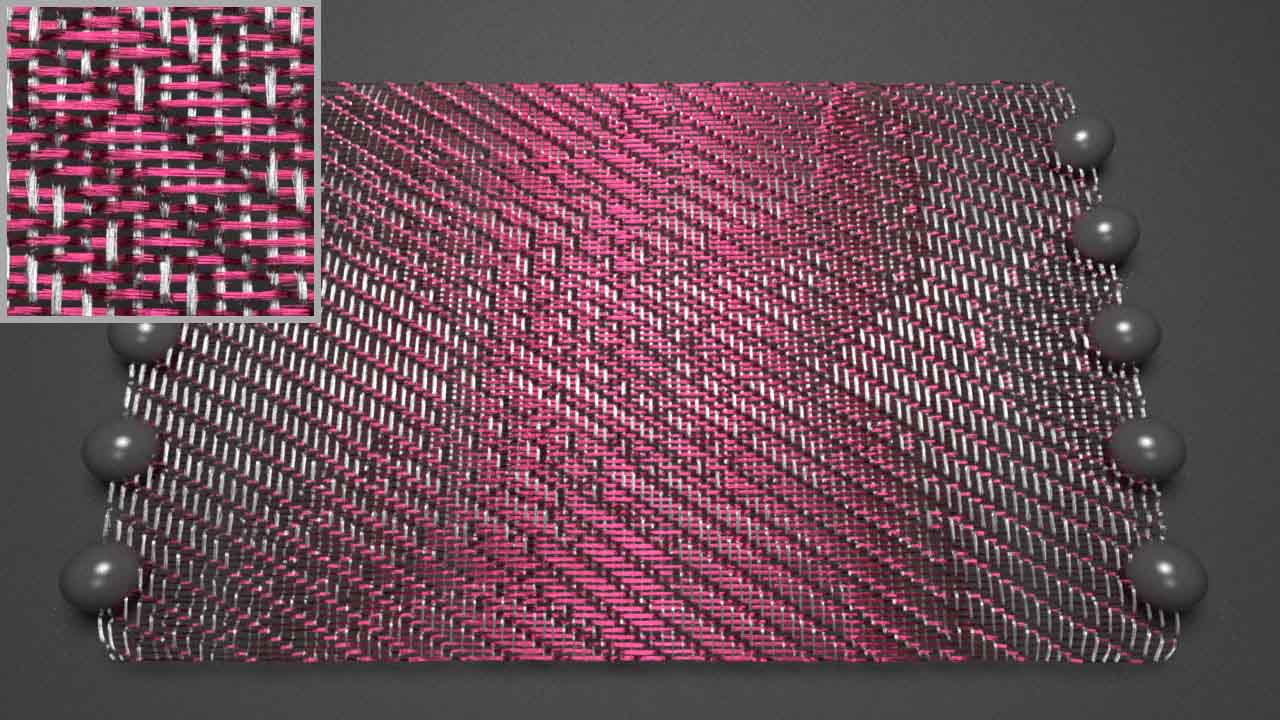}
			\put(2, 3){\small \color{white} \bfseries (b3)}
		\end{overpic}
		\\[5pt]
		\begin{overpic}[height=\resLen]{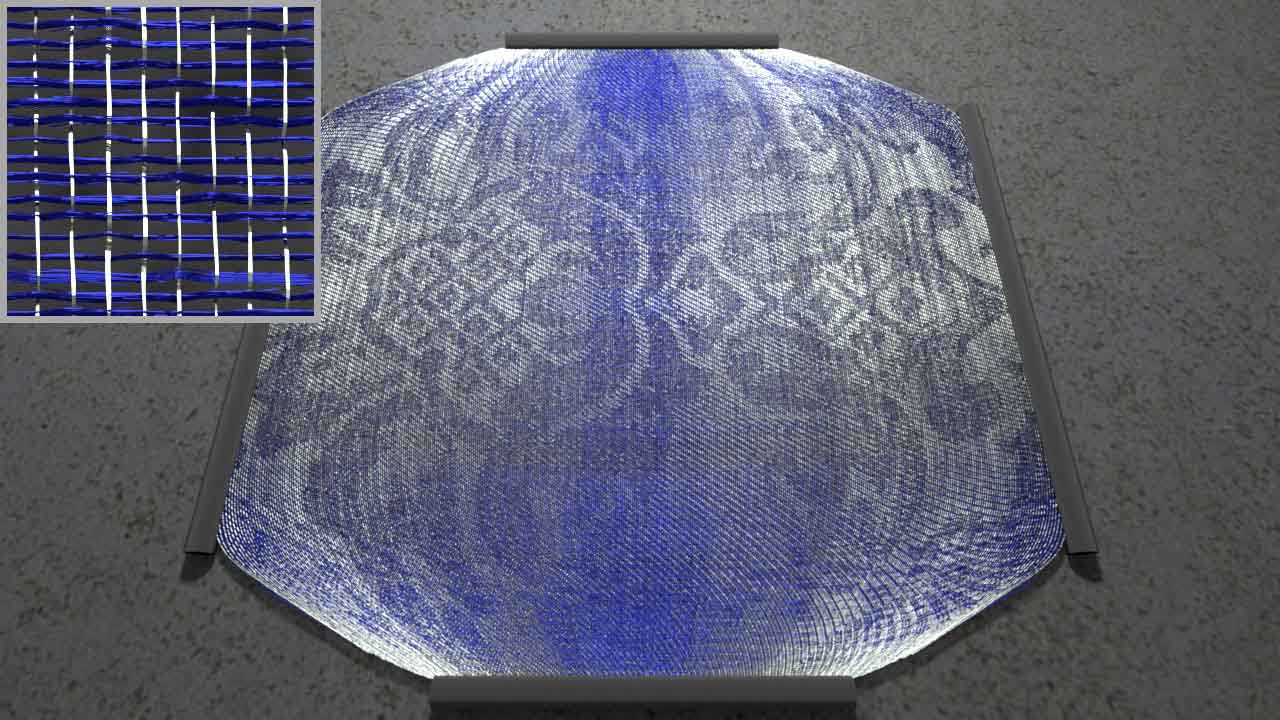}
			\put(2, 3){\small \color{white} \bfseries (c2)}
		\end{overpic}
		&
		\begin{overpic}[height=\resLen]{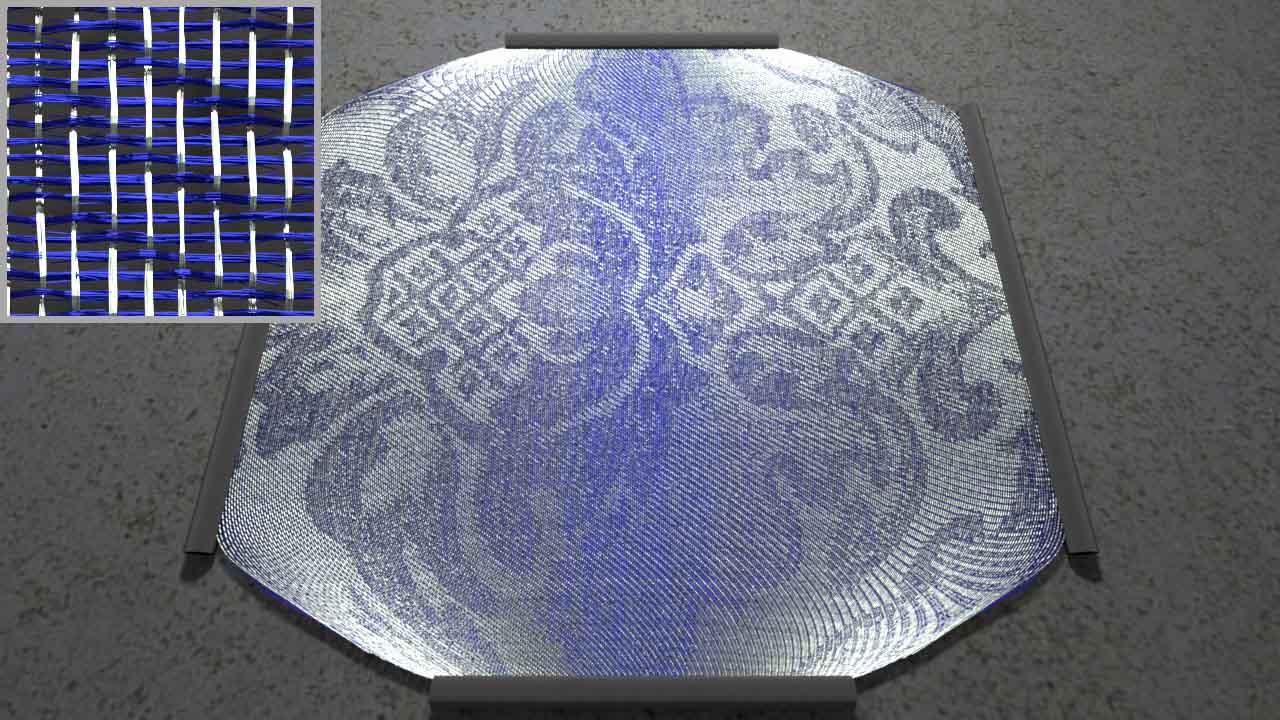}
			\put(2, 3){\color{white} \bfseries (c3)}
		\end{overpic}
	\end{tabular}
    \caption{\label{fig:result_woven}
	\textbf{Generated models for woven fabrics:}
	when applied to full fabrics, our technique is able to capture the change of appearance, both dramatic and subtle, due to fiber mechanics at both macro and micro scales.
	(1)~fabrics in their rest shapes;
	(2)~deformed fabrics modeled with our techniques;
	(3)~deformed fabrics with identical yarn centerlines but with fiber-level deformations neglected.
	The woven fabrics in (a), (b), and (c) are comprised of cotton, polyester and rayon yarns, respectively.
}
\end{figure*}

Besides knitted fabrics, those fabricated via weaving are also ubiquitous in our daily lives.
Although woven fabrics are generally more rigid (i.e., less deformable) compared to their knitted counterparts, our technique manages to capture their mechanics-driven appearance variations.
\figref{result_woven}-a contains a small woven fabric with a flower logo resulting from its weave pattern.
After being stretched horizontally, the weft yarns that follow the stretching direction are thinned more heavily than the warp yarns that are perpendicular to the stretching.
This causes the entire fabric to appear less red and more white.
Our technique successfully captures this effect (\figref{result_woven}-a2).
In \figref{result_woven}-b, we show a fabric with a $2\times 2$ twill pattern. 
This fabric is being pushed onto a static body with a capsule shape which causes the green yarns to stretch more than the yellow
yarns. This change of fiber arrangement affects the fabric's overall appearance, making the regions touching the capsule to be more yellow.  Our technique successfully captures this appearance change in colors.
Lastly, in \figref{result_woven}-c, we show a jacquard fabric being stretched both horizontally and vertically, resulting in complicated yarn and fiber deformations.
Our model manages to produce spatially varying appearance changes caused by these deformations (\figref{result_woven}-c2), which are mostly absent if fiber-level deformations are neglected (\figref{result_woven}-c3).

\begin{figure}[t]
	\centering
	\setlength{\resLen}{0.7in}
	\addtolength{\tabcolsep}{-2pt}
	\begin{tabular}{ccc}
		&
		Rest
		&
		Stretched
		\\
		\raisebox{22pt}{\rotatebox[origin=c]{90}{\small \bfseries Photo}}
		&
		\begin{overpic}[height=\resLen]{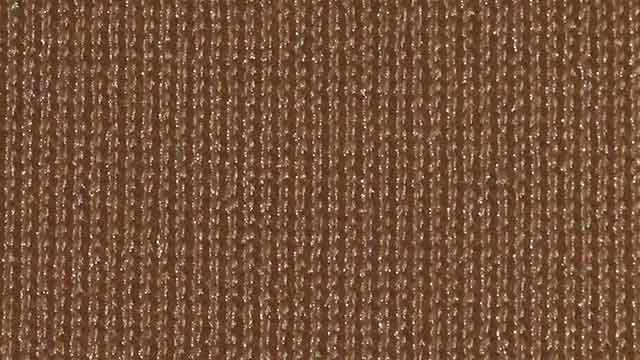}
			\put(2, 3){\color{white} \small \bfseries (a1)}
		\end{overpic}
		&
		\begin{overpic}[height=\resLen]{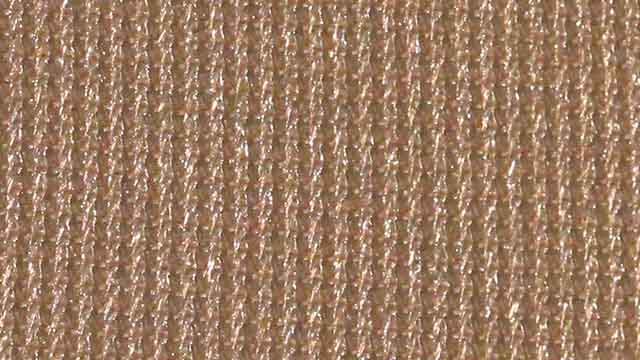}
			\put(2, 3){\color{white} \small \bfseries (a2)}
		\end{overpic}
		\\
		\raisebox{22pt}{\rotatebox[origin=c]{90}{\small \bfseries Ours}}
		&
		\begin{overpic}[height=\resLen]{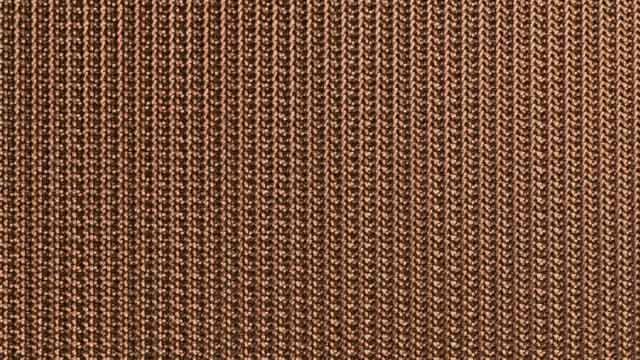}
			\put(2, 3){\color{white} \small \bfseries (b1)}
		\end{overpic}
		&
		\begin{overpic}[height=\resLen]{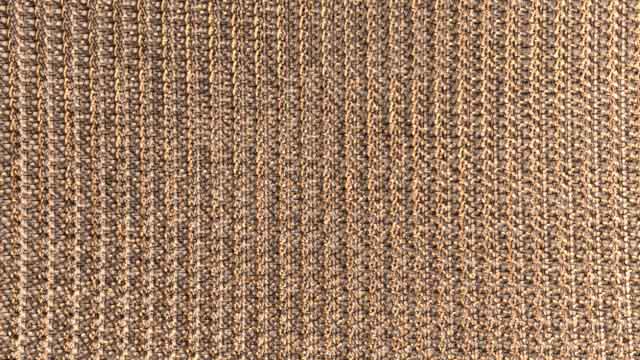}
			\put(2, 3){\color{white} \small \bfseries (b2)}
		\end{overpic}
		\\
		\raisebox{22pt}{\rotatebox[origin=c]{90}{\small \bfseries No deform.}}
		&
		\begin{overpic}[height=\resLen]{images/photoComp/simul_rest}
			\put(2, 3){\color{white} \small \bfseries (c1)}
		\end{overpic}
		&
		\begin{overpic}[height=\resLen]{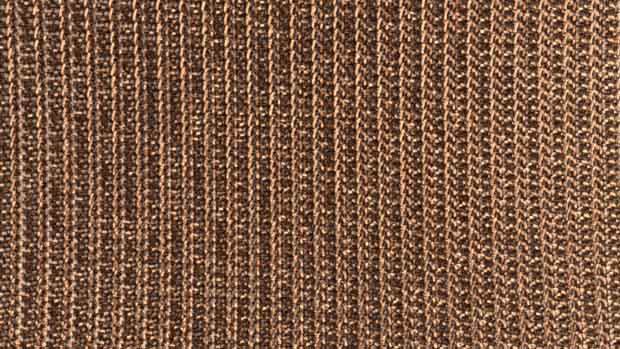}
			\put(2, 3){\color{white} \small \bfseries (c2)}
		\end{overpic}
	\end{tabular}
    \caption{\label{fig:photo_comp} 
    	\textbf{Qualitative validation} of our technique.
    	We compare photographs of a real knitted fabric made of nylon in rest~(a1) and stretched~(a2) states.
    	With a virtual model with a similar rest appearance~(b1), our method predicts its appearance change when the fabric is stretched~(b2).
    	Existing methods neglect the mechanical responses of yarns and fibers, and thus fail to capture this phenomenon~(c1, c2).
	}
\end{figure}

\section{Discussion and Conclusion}
\label{sec:conclusion}
\paragraph*{\em Limitations and future work}
Since our technique does not simulate fibers explicitly, it cannot capture \rev{visually significant effects involving}
complex fiber-level behaviors.
For instance, brushing a yarn can pull out many of its fibers and make the yarn hairier.
Capturing effects like this may require explicit simulation of individual fibers.
While currently intractable, this problem remains interesting for future investigation.
The same computational challenge also arises even in yarn-level simulation when the number of yarns are large.
Our pipeline does not critically depend on the particular simulation model that we choose to use.
Thus, more efficient yarn-level simulation methods will likely be beneficial for our pipeline. 

Further, since yarns in our yarn-level simulation have no explicit ``boundaries'', the collision is resolved using volumetric elastic energies.
Yet two yarn curves can get very close to each other, causing intersecting fibers. 
While how to resolve collisions of filament structures in a robust and efficient way remains a chronic challenge in general, improving our technique \rev{to} reduce fiber intersections is an interesting topic for future research.

\rev{Lastly, the procedural model used by our method was previously tested mainly for simple yarns~\cite{schroder2015image,Zhao:2016:FPY}.
Generalizing the model and our method to support a wider range of yarns (e.g., artificial and novelty ones) can benefit future applications.}

\paragraph*{\em Conclusion}
We introduce the first mechanics-aware cloth appearance model capable of capturing the rearrangement of cloth yarns and fibers caused by external forces without simulating individual fibers explicitly.
At the core of our technique is an extended modeling procedure that synthesizes cloth fibers and rearrange them based on simulated information only at the yarn level.
We leverage a custom-design regression neural network that maps the simulated forces expressed as deformation gradients to cross-sectional affine transformations of fiber centers.
To train this network, we use training simulations of a single yarn at both the yarn and the fiber levels and utilize a novel parameter fitting step to ensure the consistency at both scales.
Results rendered with our technique demonstrate physically plausible cloth appearance at both micro and macro scales.

\section*{Acknowledgments}
We thank the anonymous reviewers for their constructive suggestions and comments. This work was supported in part by NSF grants 1453101, 1717178, and 1813553.

%
\bibliographystyle{abbrv}
\bibliography{fibersim}
%
%
\begin{IEEEbiography}
	[{\includegraphics[width=1in,clip,keepaspectratio]{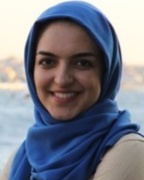}}]{Zahra Montazeri}
	is a Ph.D. candidate in the
	Department of Computer Science at University
	of California, Irvine. She received her M.Sc. degree from University
	of California, Irvine in 2017, and her B.Sc degree from Sharif University of Technology in 2015. Her research focuses on appearance modeling for complex materials such as cloth and fabrics.
\end{IEEEbiography}
\begin{IEEEbiography}
	[{\includegraphics[width=1in,clip,keepaspectratio]{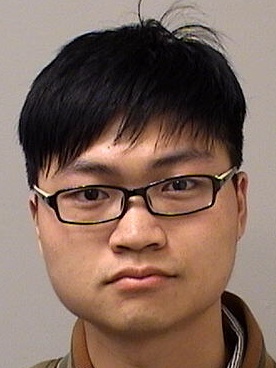}}]{Chang Xiao}
	is a Ph.D. student in Columbia University. He received the B.E. degree in Computer Science at Zhejiang University, China in 2016. His research interests are computer graphics, computer vision, and machine learning.
\end{IEEEbiography}
\begin{IEEEbiography}
    [{\includegraphics[width=1in,clip,keepaspectratio]{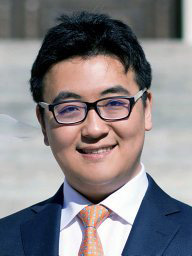}}]{Yun (Raymond) Fei} received his Ph.D. degree from Columbia University in 2019, and his B.Eng degree from Tsinghua University in 2013. As of 2019, his research focus on physics simulation of various materials, including but not limited to, hairs, clothes, liquid, sound, and the interactions between them.
\end{IEEEbiography}
\begin{IEEEbiography}
	[{\includegraphics[width=1in,clip,keepaspectratio]{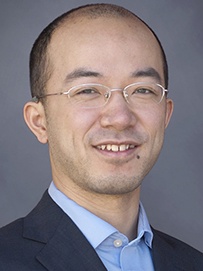}}]{Changxi Zheng}
	is an Associate Professor in
	the Computer Science Department at Columbia
	University. Prior to joining Columbia, he received
	his M.S. and Ph.D. from Cornell University and
	his B.S. from Shanghai Jiaotong University. His
	research spans computer graphics, physicallybased simulation, computational design, computational acoustics, scientific computing and robotics. He won the NSF CAREER Award and the Cornell
	CS Best Dissertation award in 2012.
\end{IEEEbiography}
\begin{IEEEbiography}
	[{\includegraphics[width=1in,clip,keepaspectratio]{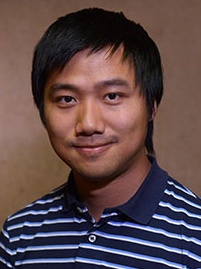}}]{Shuang Zhao}
	is an Assistant Professor in the
	Department of Computer Science at University
	of California, Irvine. Before joining UCI, he was
	a postdoctoral associate at MIT. Shuang received his M.S. and Ph.D. in computer science
	from Cornell University. His research focuses on physics-based simulation of light transport, material appearance modeling/acquisition, and physically based rendering.
\end{IEEEbiography}

\end{document}